\documentclass[11pt,twoside]{pnas-new}
\pdfoutput=1

\templatetype{pnasmathematics} 

\title{Process-guidance improves predictive performance of neural networks for carbon turnover in ecosystems}

\author[a,1]{Marieke Wesselkamp} 
\author[a,b,1]{Niklas Moser}
\author[c]{Maria Kalweit}
\author[c, d]{Joschka Boedecker}
\author[a]{Carsten F. Dormann}

\affil[a]{Biometry and Environmental System Analysis, University of Freiburg, Germany}
\affil[b]{Biological and Environmental Science, University of Jyväskylä, Finland}
\affil[c]{Department of Computer Sciences, University of Freiburg, Germany}
\affil[d]{Cluster of Excellence BrainLinks-BrainTools, Freiburg, Germany}

\leadauthor{Wesselkamp, Moser} 

\significancestatement{Deep-learning is the state-of-the-art for data-driven model predictions. Given the low sample size typical in many environmental research fields, these approaches can rarely be applied. When data are complemented by process understanding, as coded in physical or empirical models, both predictions and generalisations can be substantially improved over both data-only neural networks and mechanism-only process models. Comparing different approaches of such process-guidance helps decide on how to best combine process models and neural networks.}

\authorcontributions{CFD, NM and MW conceived the study. NM and MW coded the  neural network-process model combinations and ran the fitting. CFD, NM and MW analysed the results and wrote the manuscript.}
\authordeclaration{The authors declare no competing interests.}
\equalauthors{\textsuperscript{1}NM and MW contributed equally to this work.}
\correspondingauthor{\textsuperscript{2}To whom correspondence should be addressed. E-mail: \nolinkurl{marieke.wesselkamp@biom.uni-freiburg.de}}


\begin{abstract}

Despite deep-learning being state-of-the-art for data-driven model predictions, it has not yet found frequent application in ecology. Given the low sample size typical in many environmental research fields, the default choice for the modelling of ecosystems and its functions remain process-based models. The process understanding coded in these models complements the sparse data and neural networks can detect hidden dynamics even in noisy data. Embedding the process model in the neural network adds information to learn from, improving interpretability and predictive performance of the combined model towards the data-only neural networks and the mechanism-only process model. At the example of carbon fluxes in forest ecosystems, we compare different approaches of guiding a neural network towards process model theory. Evaluation of the results under four classical prediction scenarios supports decision-making on the appropriate choice of a process-guided neural network. 

\end{abstract}

\dates{This manuscript was compiled on \today}

\begin{document}

\maketitle
\thispagestyle{firststyle}
\ifthenelse{\boolean{shortarticle}}{\ifthenelse{\boolean{singlecolumn}}{\abscontentformatted}{\abscontent}}{}


\dropcap{T}he current revolution in deep learning has left much of environmental science unimpressed. While descriptive global studies benefit from an unprecedented wealth of data, and hence relish the opportunities brought about by the flexibilities of neural network-based analyses \cite{Asner2020, Brandt2020, Duporge2021, Jarriel2021}, local and regional studies aiming at understanding processes in an ecosystem do not. One reason is that only few ecosystem processes can be monitored, sensed and recorded at high resolution. Hence big data are comparatively rare, and deep learning provides no or only very limited improvements over traditional statistical methods and machine learning for small to moderately sized data sets \cite[e.g.][]{Bury2021}. For example, satellite imagery cannot currently sense leaf-level photosynthesis and the resulting carbon fluxes at the scale of minutes; virtually all biodiversity data, even if collected automatically, must be processed by experts to derive species abundances and richness information; data on soil processes are extremely heterogeneous, with currently hardly any chance of representation in continuously sensed data. As a consequence, the unquestionable potential of deep-learning also for environmental science \cite{Reichstein2019} can currently rarely be leveraged for prediction, or for generalising ecosystem understanding.

\begin{figure}
\centering
\includegraphics[scale=0.58]{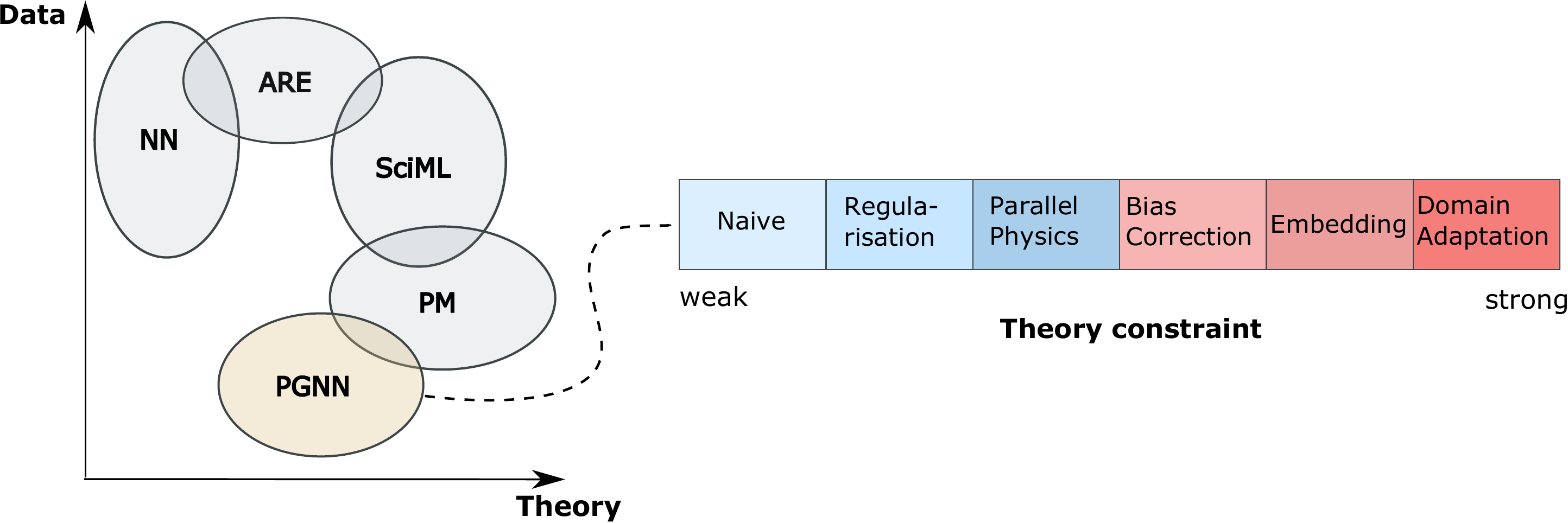}
\caption{A figure depicting the data-theory plane. 
Coloured areas reflect sweet spot of the model types in data-theory gradient. ARE = automatic reverse engineering, SciML = scientific machine learning aka NNinPM, PM = process-based models, PGNN = physics-guided neural networks.}\label{fig:side2}
\end{figure}

The relationship between data and theory in environmental modelling spans a gradient from data-driven, theory-free approaches to process-models run without link to local data (Fig.~\ref{fig:side2}) \cite{Dormann2012}. In between, four main strategies of linking data and models are emerging, of which two require substantial amounts of data: 
(i) automatic reverse engineering underlying process laws, where mathematical functions are combined optimally to describe observations  \cite{Bongard2007, Schmidt2013}; and 
(ii) partially-specified process models, with neural networks providing data-driven submodels in an otherwise process-based model \cite{Wood2001, Rackauckas2021}.

In contrast, (iii) data-calibrated process models, typically employing Bayesian or generalised Kalman-filter approaches \cite{Clark2006, Hartig2012, Dietze2017, Fer2018}, benefit from any amount of data to compare process description to, but they suffer from the need to re-run the model hundreds to many thousands of times, with only limited runtime improvement from parallelisation \cite{Speich2021}.



\begin{figure}
\centering
\includegraphics[width=\textwidth, clip]{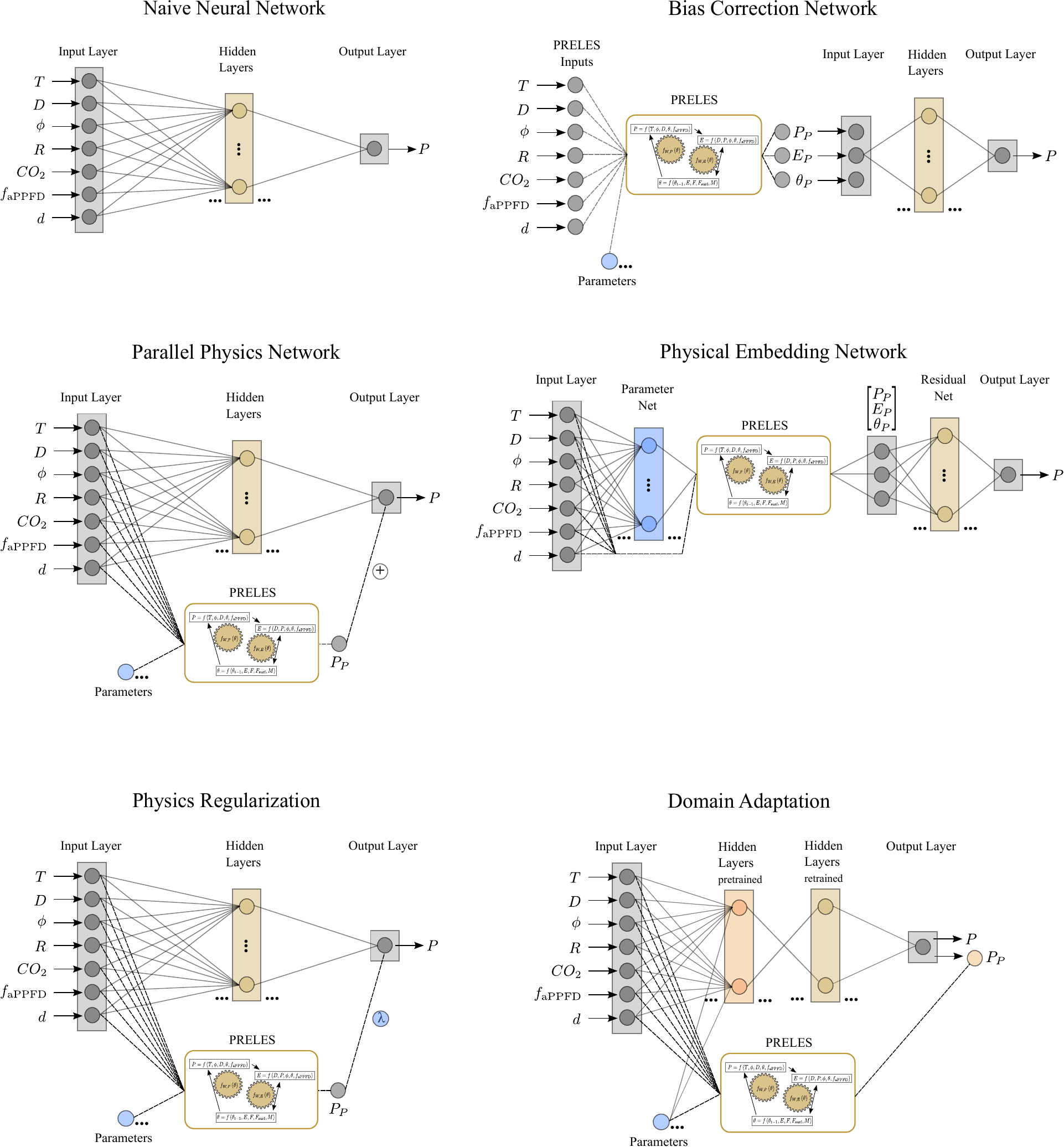} 
\caption{Schematics of the five PGNN approaches. Blue elements mark parameters of the process model PRELES, grey mark data (input, output, simulated values). A neural network is given for comparison. The figure for ``process regularisation'' is identical to that of ``residual physics'', except that the two predictions are not added but evaluated as a joint likelihood (see text). Ellipses (...) indicate more layers and nodes, according to hyper-parameter search. $P$ refers to prediction made by the network to be compared with observations.}
\label{fig:PGNNs}
\end{figure}

Process-guided neural networks (iv), where process models are embedded, in one way or another, into a deep-learning pipeline \cite[e.g.][]{Ba2019, Karpatne2018, Zhao2019}, seem to be particularly well-poised to be using the relatively sparse data in environmental science. 
Moreover, they represent general structures in the data that remain unexplained in current theories, thus provide higher generalisability than the two separate approaches and indicate yet undetected processes.  
Process-guided neural networks use theory and empirical knowledge, as coded into the process model, to constrain the network's flexibility when fitting it to observations. Within this group of approaches, we can differentiate between five designs, although the terminology is in constant flux (Fig.~\ref{fig:PGNNs}):

\begin{enumerate}
    \item Domain adaptation (aka transfer learning, aka physics fusion, aka pre-training). The process model is used to simulate many data sets akin to those observed, but within the range of expected parameter values. A neural network is then trained to these big-data simulations; only the last layers of this pre-trained network are then re-fit to the actual observed data.
    \item Residual correction (aka bias correction). A neural network is trained on the observed data using as input the model output. It thereby aims at removing any bias inherent in the process model output.
    \item Parallel physics (aka residual physics). A neural network is trained in parallel to the immutable process model, and their predictions are averaged before fitting to the data. As a result, the network learns consistent deviations between data and the process model, trying to link process-residuals to input variables.
    \item Physics regularisation. The network is trained using a loss function that consists of the cross-validation based weighted sum of distance to  data and  distance to the process-model simulations. The data can therefore not drive the network too far away from the process model.
    \item Physics embedding. The model parameters are turned into functions of the input predictors, by having a network linking input to parameters, and a second network provides additional bias correction. The entire setup is trained as a deep-learning model.
\end{enumerate}

The magnitude to which these approaches' flexibility in the training procedure are constrained by the embedded theory vary from very weak through data structure only to very strong through prior parametric assumptions \citep{karniadakis2021physics}. To date, these process-guided neural network approach have not been compared for their on-site predictive performance, nor for their ability to generalised beyond the training site. In this study, we evaluate predictive and transfer performance, comparing the PGNN approaches to a process model fitted to local data, and to a multi-layer neural network, representing the two extremes of the data-theory-gradient. We expect the multi-layer neural network with the highest flexibility in fitting the data to show the best predictive performance and the constraints in the PGNNs to prevent unrealistic predictions.
As a case study, we use carbon and water fluxes in forest ecosystems. 


\section*{Methods}

\subsection{Study sites and data}
Five forest sites are chosen from the PROFOUND database based on their data availability of environmental covariates and carbon turnover  \citep{reyer2020profound}. The forest sites cover dominantly decidious, mixed and coniferous forest types with varying dominating species (see Supplementary Information). The environmental covariates of the forest sites represent conditions of mediterranian, temperate and boreal climates.    

The PROFOUND database provides time series data in two resolutions. Eddy-covariance measurements from FLUXNET are recorded on a daily or half-hourly resolution. Satellite imagery data is provided from MODIS on a 8-day resolution. We choose variables and their resolution based on what is commonly used in modelling approaches with the process model. 

Carbon turnover is measured as gross primary productivity (GPP, $P$). GPP measurements of half-hourly resolution are taken from FLUXNET and aggregated to daily means. 
The environmental covariates are air temperature ($T$), precipitation above canopy ($R$), photosynthetic active radiation ($\phi$) and the fraction of absorbed photosynthetic active radiation ($f_{\text{aPPFD}}$). They are taken from FLUXNET in daily resolution. The fraction of absorbed photosynthetic active radiation is taken from MODIS in 8-day resolution and converted to daily resolution by assuming constant values within the 8-day periods.

The data set was aligned in terms of resolution and units in a preprocessing procedure that is described in the Supplementary Information.
The resulting dataset consists of daily measurements of covariates and carbon turnover for five forest sites over four years.

We denote the general modelling exercise in this analysis by the mapping $f: X \mapsto Y$, where $X$ is the $k$-th row vector of the environmental covariate matrix $X = \lbrace T_k, D_k, \phi_k, R_k, f_{\text{aPPFD}, k} \rbrace$ and $Y$ the $k$-th entry of the GPP vector $Y = \lbrace P_k \rbrace$. 
\subsection{Process model}\label{sec:process model}

The process model used to predict carbon and water fluxes in this case study is Preles (PREdict Light-use efficiency, Evapotransipration and Soil water content) \cite{peltoniemi2015semi}. Preles is a semi-empirical model that predicts gross primary productivity ($P$), evapotranspiration ($E$) and soil water content ($\theta$) 
on a daily time scale. More specifically, PRELES does the mapping $f:X \mapsto Y$
with $X =  \lbrace T, D, \phi, R, f_{\text{aPPFD}}, CO_2, d \rbrace$ and $Y = \lbrace P, E, \theta \rbrace$, where $d$ is the day of the year and $CO_2$ is a constant denoting the carbondioxide concentration (in ppm). 

The mapping $f$ consists of three subsystems for $P$, $E$ and $\theta$ that are linked to each other. The calculation of $\theta$ is based on a three pool formulation, splitting $\theta$ into soil, surface and snow pools. The state of $\theta_k$ at day $k$ depends on $\theta_{k-1}$ the state of the previous day. 


The subsystem for $P$ prediction on a day $k$ is based on a general model for light-use efficiency \cite{makela2008developing}:
\begin{equation}
    P_k = \beta \phi_k f_{\text{aPPFD}} \prod_{i} f_{i,k} \quad ,
\end{equation}
with $\beta$ as the potential light-use efficiency and  $\prod_{i} f_{i,k}$ the product of exogenous driven modifiers for $\beta$. One of these modifiers, the soil water modifier $f_{W,P}$ links the subsystems of $\theta$ and $P$.

In the third subsystem $E$ is calculated as directly depending on $P$. The subsystem is connected to $\theta$ by a modifier $f_{W,E}$.
For more details on the process model structure see \cite{peltoniemi2015semi}.


\subsection{Neural network}
Neural networks do the mapping $f: X\mapsto Y$ through a hierarchically structured latent feature space (\textit{hidden layers}). The latent space allows for the representation of hidden processes alongside the measured input $X$. Through layer-specific transformations (\textit{activation functions}), the approximated function $f$ can be highly non-linear.
The output of a neural network with a single hidden layer is defined as
\begin{equation}
y_j = \sigma \left( \left(\sum_i x_i \times W_{ji} \right)  + b_j \right) \quad \text{,} 
\end{equation}
where $y = \lbrace y_1,..., y_j \rbrace \in Y$ is the $j$-dimensional output vector and $x = \lbrace x_1,...,x_i \rbrace \in X$ is the $i$-dimensional input vector. The weight matrix $W_{ji}$ and the bias vector $b_j$ are network parameters that are learned during the training procedure. The linear combination of inputs, weights and bias is transformed by the activation function $\sigma$ which is part of the hyperparameters of the network.

Here, a most basic neural network type is used that is a multi-layer perceptron (MLP). Through the choice of the linear activation function in the last layer and the mean squared error as the objective function, the MLP performs a regression task under the assumption of normality. 
Used without process-guidance, this MLP is referred to as the naive neural network.

\subsection{Process-guided neural networks}\label{sec:pgnns}

\subsubsection{Types of process-guided neural networks}

\paragraph{Residual correction}
The residual correction is sequentially structured in two parts (Fig.~\ref{fig:PGNNs}). First, information flows forward-only through the PM. The PM uses the same inputs and parameters as the stand-alone PM. Second, the PM predictions for GPP, evapotranspiration and soil water content are used as inputs for a fully-connected feed-forward neural network. The neural network maps $f:X\mapsto Y$, where $X = \lbrace P_p, E_p, \theta_P \rbrace$ is the PM predictions vector for GPP, evapotranspiration and soil water and $Y= \lbrace P \rbrace$ is the GPP ouput vector. 
In the training procedure, the weights and biases of the neural network are adjusted based on the loss function
\begin{equation}
    \mathcal{L} = \mathcal{L}_{\text{MSE}}(y, \hat{y}) \quad \text{,}
\end{equation}
where $\hat{y}$ are the predicted GPPs of the neural network. 

\paragraph{Parallel physics}
The parallel physics parallelises forward information flow through the PM and a fully-connected feed-forward neural network simultaneously (Fig.~\ref{fig:PGNNs}). The PM and the neural network use the same input variables. Information flows only forward through the PM. The PM uses the same parameter values as in the stand-alone approach. The predicted GPP of the PM and the neural network are summed. The neural network maps $f:X\mapsto Y$ with $X = \lbrace T, D, \phi, R, f_{\text{aPPFD}}, d \rbrace$ and $Y = P = \lbrace P_{\text{NN}} + P_{\text{PHY}} \rbrace$ where the GPP prediction $P$ is the sum of the neural network prediction $P_{\text{NN}}$ and the PM prediction $P_{\text{PHY}}$. 
Hence, the loss $\mathcal{L}$ is calculated using the mean squared error of GPP predicted by the PM $\hat{y}_\text{phy}$,  the neural network predictions $\hat{y}_\text{nn}$ and observations $y$ as
 \begin{equation}
     	\mathcal{L} = \mathcal{L}_{\text{MSE}}\left(y, \hat{y}_{\text{nn}} + \hat{y}_{\text{phy}}\right) \quad \text{.}
 \end{equation}
 
The information of the mismatch between the summed predictions and observations is used to adjust the weights and biases of the neural network. 

\paragraph{Physics regularisation}
The physics regularisation is structurally similar to the parallel physics approach (Fig.~\ref{fig:PGNNs}). The neural network maps $f:X\mapsto Y$ with $X = \lbrace T, D, \phi, R, f_{\text{aPPFD}}, d \rbrace$ and $Y = \lbrace P \rbrace$.
The predictions of the PM are only used in the training procedure as part of the loss function, such that the fitted neural network minimises deviation from observation while maintaining also minimal distance to PM predictions. This prevents fits to deviate substantially from known processes.
The loss $\mathcal{L}$ is calculated with the neural network predictions $\hat y_\text{nn}$, the PM predictions $\hat{y}_\text{phy}$ and the observed GPP $y$
as
\begin{equation}\label{eq: RL}
	\mathcal{L} =  \mathcal{L}_{\text{MSE}}\left(y, \hat{y}_{\text{nn}}\right) + \lambda\  \mathcal{L}_{\text{MSE}}\left(\hat{y}_{\text{phy}}, \hat{y}_{\text{nn}}\right) \quad \text{,}
\end{equation} 
with $\lambda \in  (0, 1]$ determining the strength of the regularisation and being defined in the hyperparameter search.


 \paragraph{Physics embedding}
 The physics embedding merges the PM into the neural network modelling framework. The network comprises three modelling exercises (Fig.~\ref{fig:PGNNs}). First, the parameter net maps environmental covariates to the PM parameters, $f:X\mapsto Y$ with $X = \lbrace T, D, \phi, R, f_{\text{aPPFD}}, d \rbrace$ and $Y = \lbrace \theta_{\text{PHY,1}},..., \theta_{\text{PHY,13}}\rbrace$. 
 Second, the PM is run with the parameter estimates $\theta_{\text{PHY}}$ from the parameter net and the environmental covariates $X$.
 Third, the PM predictions $\lbrace P_P, E_P, \theta_P \rbrace$ are transformed in the residual net into GPP predictions. The residual net is akin to the residual physics setup above. The loss function is the same for physical regularisation. Based on the loss, both the parameter net and the residual net are optimised simultaneously. 
 
 Technically, the PM is defined as a certain type of forward pass connecting only the output layer of the parameter net and the input layer of the residual net. To guarantee that during the training the information of the gradients flows backwards through the PM, it needs to satisfy tensor compatibility. In our approach, every elementary operation in the PM is therefore changed to its Torch tensor equivalent so that for each forward pass of a tensor every operation on it is tracked and stored together with its gradient. During backpropagation the information of the gradients is back-traced through the stored computation graph of the tensor.  

\paragraph{Domain adaptation} 
The domain adaptation consists of a pre-training on data simulated by the process model, and a re-fitting of the resulting network to the actual observation \cite{Gongora2019}. 
In our case, the simulated dataset is purely artificial and generated by prediction from the PM, based on simulated weather data and parameter samples from their prior distributions. The architecture and hyper-parameters for the domain adaptation network are optimised for the observations and as such equal to the naive neural network. Without using an encoder network, the domain adaptation network requires homogeneous input dimensions. We take only environmental covariates as explicit input in pre-training and neglect the parameter prior samples under the assumption that the same information is represented in the simulation from the process model. 

The key challenge for domain adaptation is to choose a restricted but realistic set of simulation parameters and environmental covariates: too wide, and the pre-training will not restrict the resulting network; too narrow, and the fitted network will not be able to generalise to conditions outside the pre-training range. Technical details on the simulated input data can be found in the Supplementary Information.

\subsection{Training and test procedure}

\paragraph{Neural Network} Due to the parametric capacity of neural networks, a data splitting into training and validation set is already performed during the parameterisation procedure (\textit{training}). Furthermore, one year of the data is hold out as test set until evaluation. 
Training is done in a temporal-blocked cross-validation, in which the temporal structure is kept during splitting \citep{roberts2017cross}. In both the training and evaluation processes, splits are done such that the model can be validated on a full year. Models are trained for 5000 epochs with the Adam optimisation algorithm.

\paragraph{Process Model} The default parametrisation of the PM is adjusted for 13 parameters simultaneously. The calibration is pursued in the R framework using the Bayesian Tools package for Bayesian model calibration \citep{hartig2019bayestools}. The block-cross validation splits used for the neural network training are also used during PM calibration. The likelihood is taken as the objective function. A Markov Chain is generated with a DREAM$_\text{Z}$ sampler \cite{terBraak2008} over 50000 iterations. 

\subsubsection{Network architecture search}

Selecting the neural network architecture and the parameters of the optimisation algorithm can be understood as an optimisation of neural network hyper-parameters. These are optimised separately by two random grid searches. 
A random grid consists of randomly drawn combinations of hyper-parameters from a given base set.
First, the network architecture is selected for random parameters of the optimisation algorithm, which is then adapted to this architecture in a second optimisation.
Hyper-parameters for the architecture of the MLP are the number of hidden layers and nodes, for the optimisation algorithm the learning rate and the batch size.
Both, the embedded network and the regularisation network, have a regularisation parameter $\lambda$, which is included in the architecture random grid search.
Optimisation of the architecture and hyper-parameters is performed in a two-fold cross-validation setting.

\subsubsection{Types of prediction}

The prediction experiment is conducted under four scenarios that represent classical tasks in environmental modelling. 
Predictions are made firstly, on a single site in time and secondly, on multiple sites in space. For both the on-site and the multi-site prediction tasks, two scenarios of data availability are considered. First, the time series is used on a daily resolution, that is simply the full pre-processed data set. Second, a time series with weekly resolution is simulated through systematically thinning out of the data set to every seventh day only, leading to substantially sparser data. The calibration of the PM and training procedure of the neural networks was separately pursued at on-site or multi-site data under sparse or full data availability conditions.

\subsection{Individual conditional expectation}

An individual conditional expectation analysis (ICE) is a local method to evaluate the relationship of the dependent variable and each individual explanatory variable \citep{molnar2022}. 
The interaction of the dependent variable and one explanatory variable is examined by varying the explanatory variable of interest over a pre-defined range at fix values of the remaining variables. 
Local denotes the analysis of individual predictions for each instance that in this case are GPP predictions for every day. We pursue the analysis during four period of the test year. 
These are two weeks around four days that are taken to be seasonally representative. 
We decided on the four record days for seasonal change that are the 20th of March (spring), the 21st of June (summer), the 20th of September (autumn) and the 21st of December (winter).
The dependence of the target variable is evaluated at each day during this period. This local dependence analysis corresponds to a conditional effects analysis and, in contrast to a partial dependence analyses, allows for more insight in presence of interactions among explaining variables.


\section*{Results}
\setcounter{subsection}{0} 

\subsection{Predictive performance and extrapolation}

\begin{figure}
\centering
\begin{subfigure}[b]{0.4\textwidth}
\includegraphics[width=\textwidth]{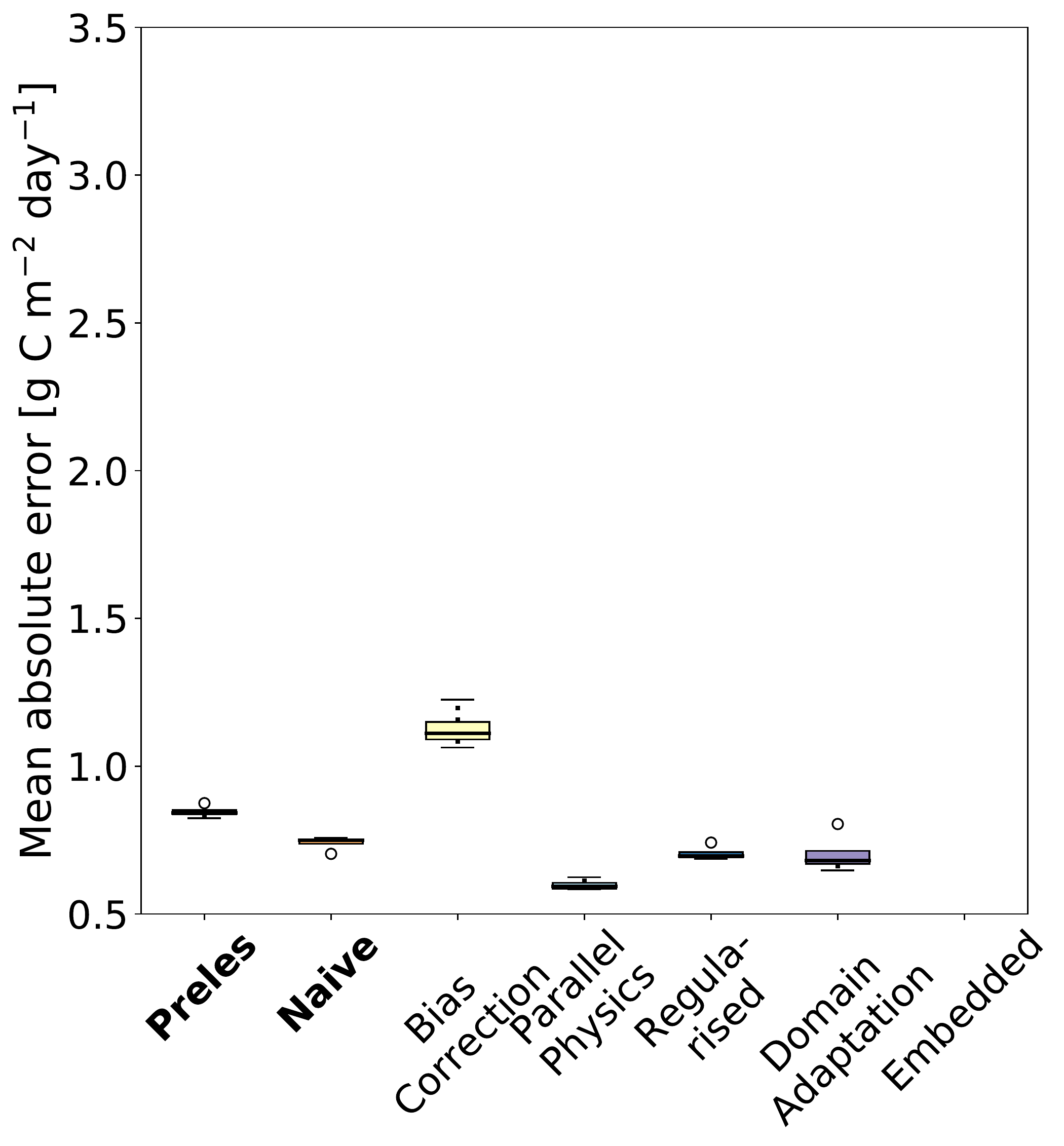}
\caption{On-site prediction with full data}
\end{subfigure}
\begin{subfigure}[b]{0.4\textwidth}
\includegraphics[width=\textwidth]{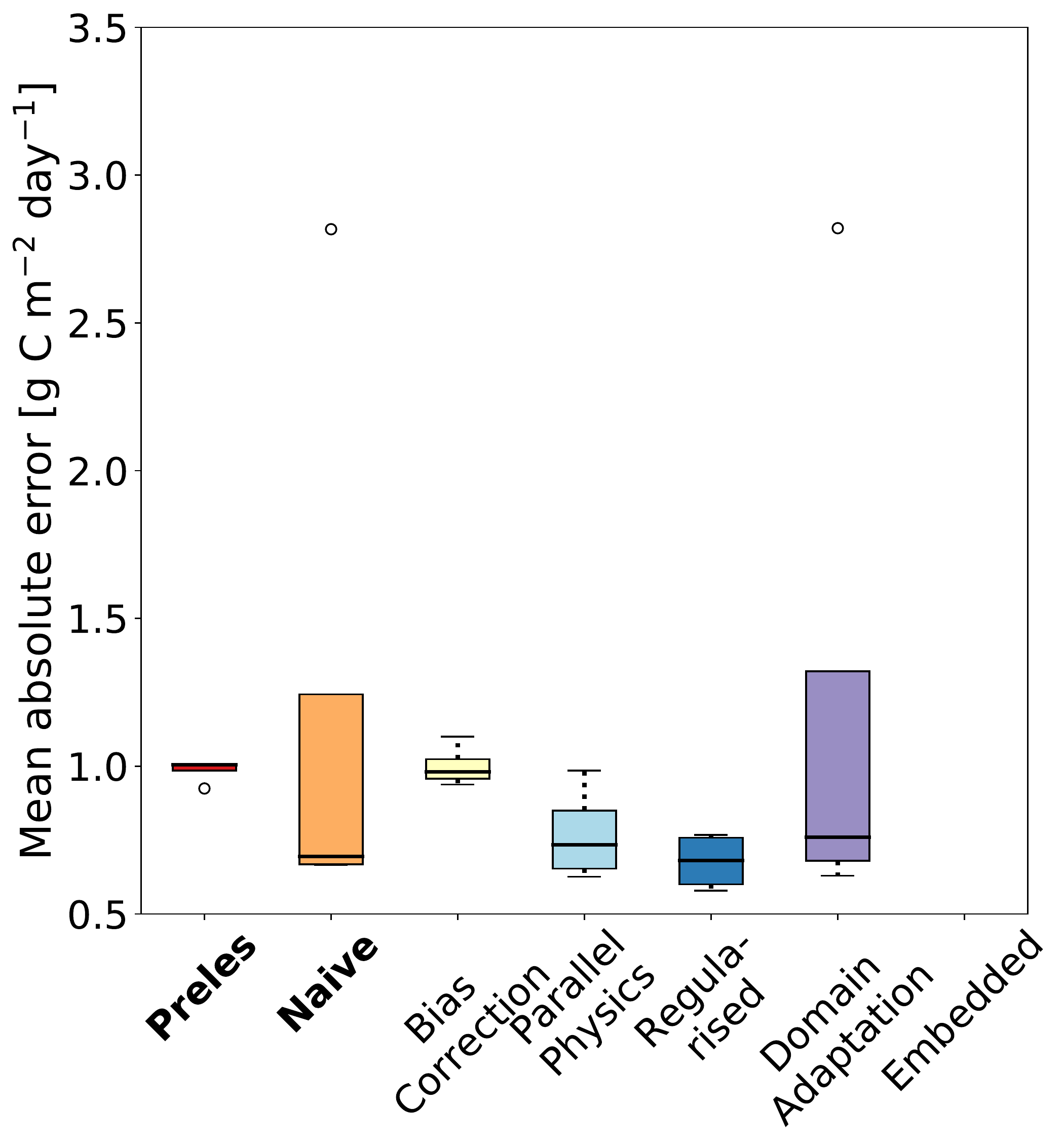}
\caption{On-site prediction with sparse data}
\end{subfigure}
\begin{subfigure}[b]{0.4\textwidth}
\includegraphics[width=\textwidth]{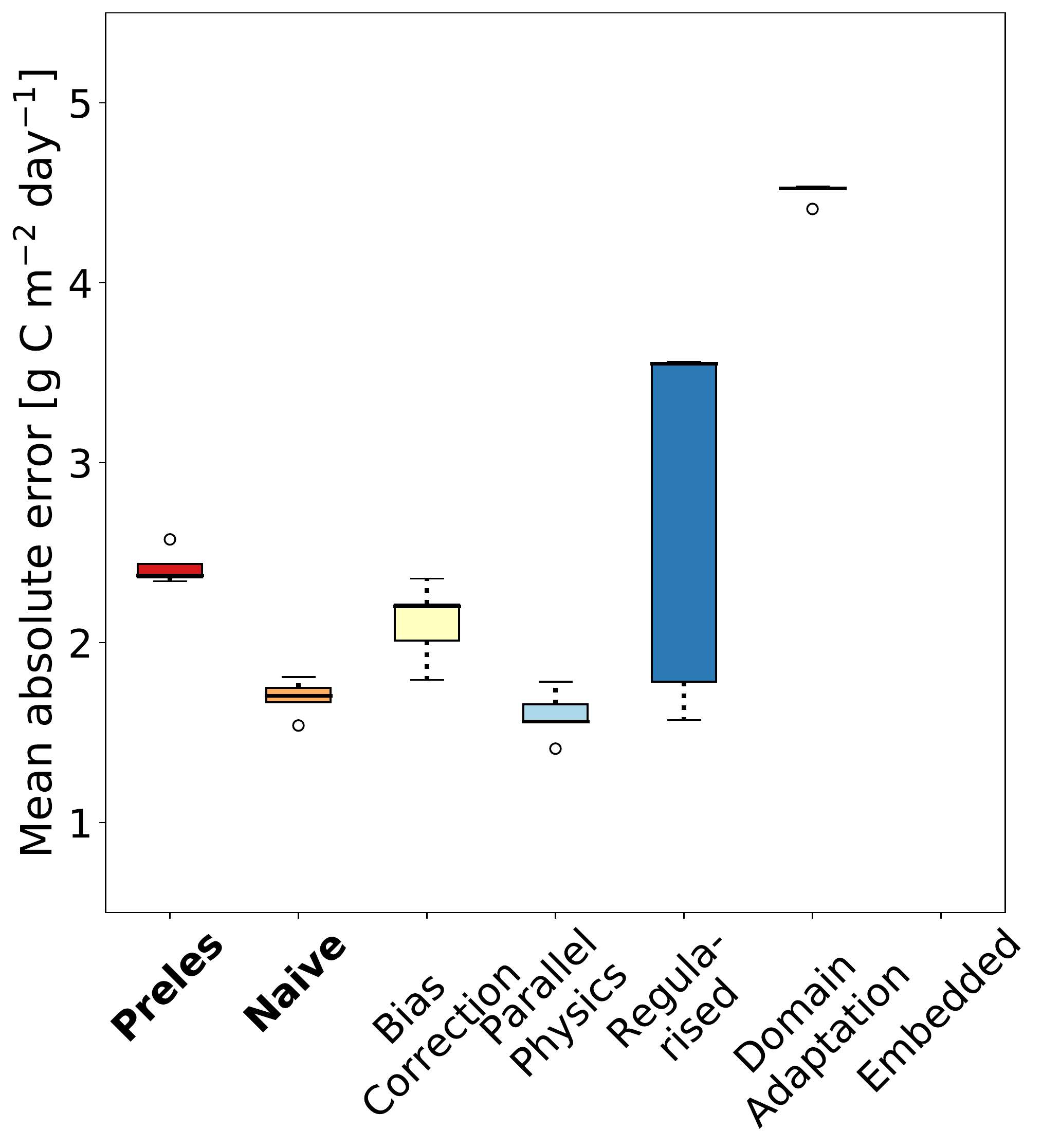}
\caption{Multi-site prediction with full data}
\end{subfigure}
\begin{subfigure}[b]{0.4\textwidth}
\includegraphics[width=\textwidth]{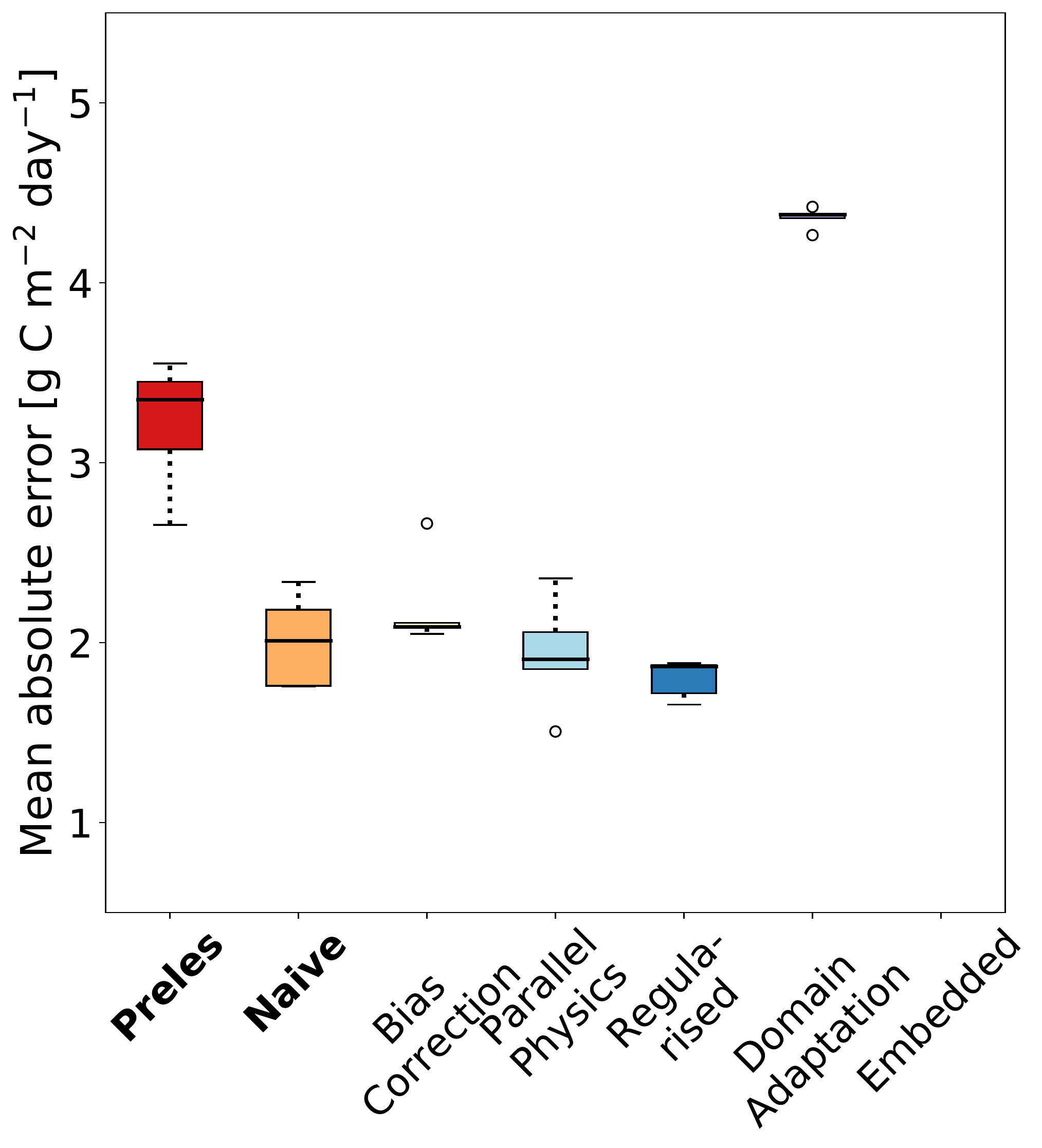}
\caption{Multi-site prediction with sparse data}
\end{subfigure}
\caption{The mean absolute error over 5000 training runs for the naive neural network and the PGNNs in the temporal (first row) and spatial (second row) prediction setting under full and sparse data scenarios (\textit{Results for the Embedded PGNN will follow}). 
}
\label{fig:performance_temporal}
\end{figure}

Both the fitted process model and a naive multi-layer perceptron (MLP) are the references for the five PGNN approaches. For the full data case, ``on-site'' validation predictions to new years by the naive network outperformed the fitted process model, but all PGNNs except bias correction provided an improvement to MLP prediction at least under one prediction scenario (Fig.~\ref{fig:performance_temporal} a, Tab.~\ref{tab:performance}). Generalising to other sites with the full data (Fig.~\ref{fig:performance_temporal} c), however, substantially increased prediction errors of all models, while improvement in prediction performance relative to that of the process model increased.

In the sparse data situation, when the data was systematically thinned out to a weekly resolution, the picture changed substantially. On-site predictions to new years was much more variable, and only parallel physics and physics regularisation improved consistently on the MLP (Fig.~\ref{fig:performance_temporal} b). Note that still all other approaches (apart from bias correction) did as well or better than the MLP, but all these approaches suffered from high variability. When predicting to other sites, the parallel physics and physics regularisation were able to improve the naive MLP predictions, the parallel physics for both, full and sparse data, the regularisation only for the latter.  Domain adaptation, which slightly improved reference model predictions worked well for on-site prediction with full data, but in the multi-site setting fared worse even than the fitted process model itself.

\begin{table}
\centering
\caption{Mean absolute validation error of GPP predictions (in gCm$^{-2}$day$^{-1}$) in a five-fold cross-validation. The best performing network under each experiment is highlighted in bold. The amount of training days varied among experiments (On-site full: 1460. On-site sparse: 207. Multi-site full: 2190. Multi-site sparse: 313).}
\begin{tabular}{lrrrrrrr}
  & \thead{Preles} & \thead{Naive} & \thead{Bias\\ Correction} & \thead{Parallel \\ Physics}   & \thead{Regularised} & \thead{Domain\\ Adaptation} & \thead{Embedding} \\
\midrule
On-site full   &  0.85$\pm$0.02 & 0.74$\pm$0.02 & 1.13$\pm$0.06 & \textbf{0.6}$\pm$0.02 & 0.7$\pm$0.02 & 0.7$\pm$0.06 & - \\
On-site sparse&  0.99$\pm$0.03 & 1.22$\pm$0.92 & 1.0$\pm$0.06 & 0.77$\pm$0.14 & \textbf{0.68}$\pm$0.08 & 1.24$\pm$0.91 & - \\
Multi-site full & 2.42$\pm$0.08 & 1.69$\pm$0.09 & 2.11$\pm$0.19 & \textbf{1.59}$\pm$0.12 & 2.8$\pm$0.92 & 4.5$\pm$0.05 & -  \\
Multi-site sparse &  3.22$\pm$0.32 & 2.01$\pm$0.23 & 2.2$\pm$0.23 & 1.94$\pm$0.28 & \textbf{1.8}$\pm$0.09 & 4.36$\pm$0.05 & - \\
\bottomrule
\end{tabular}
\label{tab:performance}
\end{table}

\subsection{Individual conditional expectations}

The ICE analysis is pursued for the most consistent performing PGNN that is the parallel physics network and results are contrasted with the ICE of the reference models, Preles and the naive neural network.
The plots show the type of relation between an explanatory variable and the target variable. 
We expect dependencies represented by the naive neural network to appear visibly more complex than those represented by the process model and the PGNN to constrain these complex dependencies to closer approximate those of the process model.

\begin{figure}
    \centering
    \includegraphics[width=.7\textwidth]{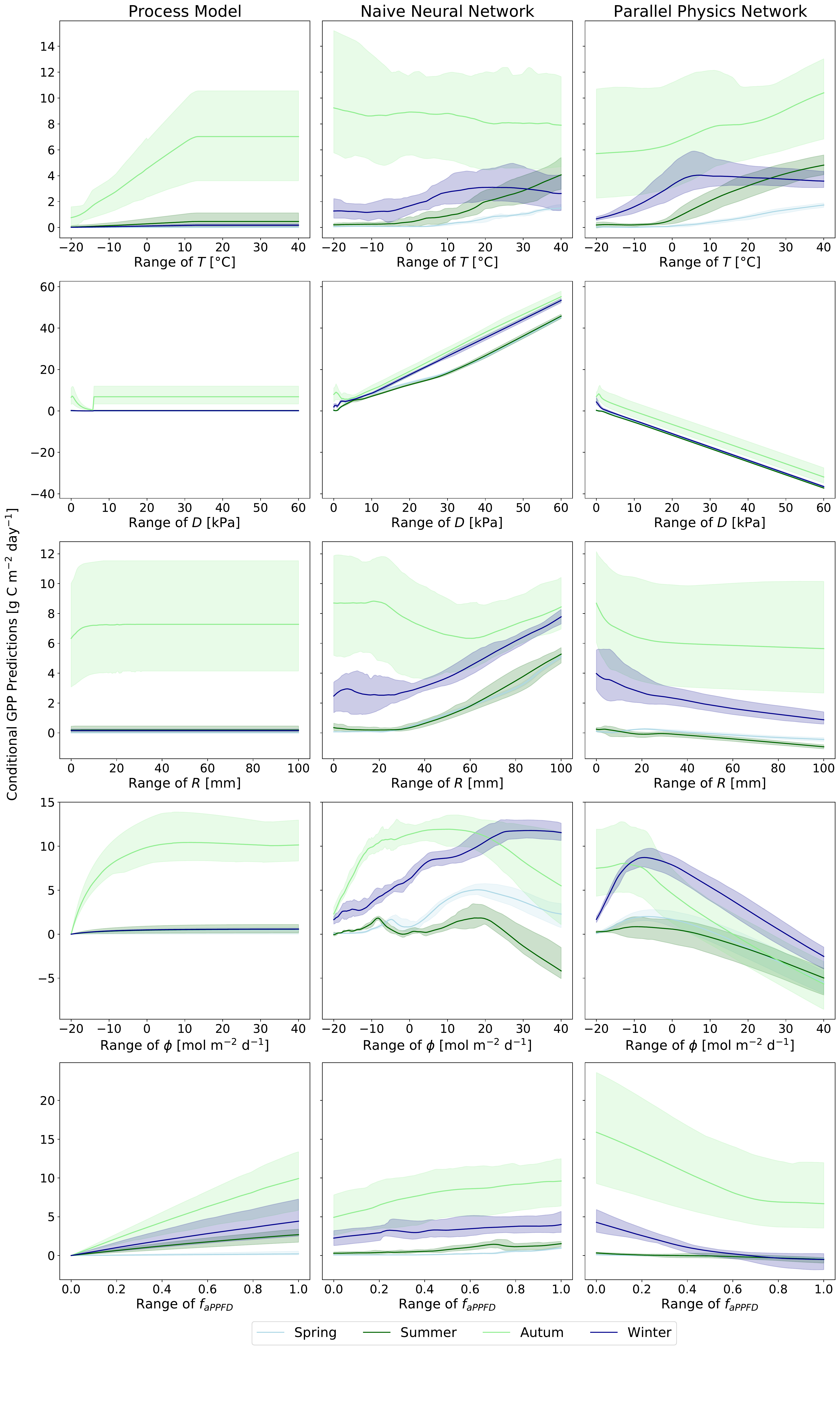}
    \caption{Individual conditional expectations of GPP predictions for each explanatory variable. Shades indicate the 95\% confidence interval during the period of two weeks around the seasonal record days.}
    \label{fig:GPPeffectsize}
\end{figure}

The effects of the explanatory variables in the process model are smooth and either quadratic ($\phi$) or linear ($f_\text{aPPFD}$)(Fig.~\ref{fig:GPPeffectsize}). 
Effects are only evident in the predictions for summer. A flat curve as in the case of $D$ indicates for the individual case that there is no effect of this variable on the prediction. 
In this, the process model differs from both the naive neural network and the parallel physics network. Except for $f_\text{aPPFD}$, all variables show a clear effect on the model prediction. 
In the case of the Parallel Physics network, where the ICE analysis is only on the input of the network, we see the effect of error correction from the naive network to the process model shown. This is clearly visible in the case of $\phi$ and $D$, where the network corrects a strong positive linear weighting in high value ranges by a negative linear weighting.

\section*{Discussion}

We compared a selection of process-guided neural networks (PGNNs), an emerging modelling approach that integrates purely data-driven knowledge with process theory.
At the example of carbon and water fluxes in forest ecosystems, the on-site predictive performance of the approaches and their ability to represent yet undetected generalisable processes in the data is evaluated. 
Exceeding our expectations, the PGNNs outperform not only to the process model in predictive accuracy, but in some cases even the multi-layer neural network. Regarding other methods that combine data and process knowledge (see Fig.~\ref{fig:side2}), the relevance of PGNNs is founded in three classical problems in process-oriented environmental modelling.
\begin{itemize}
    \item A lack of data availability, which limits options for process model calibration and the use of deep neural networks \citep{Reichstein2019}.
    \item A lack of process understanding, which manifests itself as structural mis-specification in process models \citep{Wood2001}.
    \item The complexity of ecosystems, limiting generalisability of process models.
\end{itemize}
On one hand, PGNNs turn out to be especially suitable when data is sparse or of low resolution. On the other hand, they allow considerably better predictions than the process model when data descends from different sources/sites.
An approach commonly used to analyse variable importance in statistical models and currently to interpret results of deep neural networks \citep{molnar2022} suggests that successful process-guidance is possible.

\subsection*{PGNN-specific use cases for prediction improvement}\label{sec:usecases}

The four experimental scenarios under which the approaches were evaluated demonstrate that the improved predictive performance of PGNNs towards the naive neural network and the PM is both, task- and PGNN-specific.
The appropriate use case of any of the different PGNNs depends on the prediction goal and the data availability.
The parallel physics network predicts best when data is available on a daily resolution, whereas when data is sparse, here on a weekly resolution, the regularised network outperforms it (see Fig.~\ref{fig:performance_temporal}). Both approaches integrate the physics into the neural network through representing the residuals in the loss function, eventually corresponding to a physics-based likelihood assumption (see section \ref{sec:pgnns}). However, considering residuals in an additive loss term with an own regularisation parameter enables the possibility of shrinking the importance of the PM residuals to zero, making the regularised neural network more flexible than the parallel physics network.
This highlights the varying degrees to which the PM constraints a neural network. We can assume that predictive performances of approaches that experience a hard constraint will closer approximate the predictive performance of the PM. Imagining a full constraint, we can expect the PGNN to perform at maximum as good as the PM, if not even worse. This exemplary happened when imposing a hard constraint on the training process through weight initialisation in the domain adaptation. Incorporating the PM residuals as a physics-based likelihood in contrast constitutes a soft penalty constraint on the loss function, allowing for more flexibility in the remaining loss term \citep{karniadakis2021physics}.

\subsection*{Generalisability as process-guidance}

The most simple and natural constraint to neural networks in terms of process-guidance descends from the observational data itself.
The neural network is forced to represent the detectable underlying data generation processes, which can be seen as a integration of processes \citep{karniadakis2021physics}. This is a weaker constraint than a soft penalty constraint on the loss function and constitutes the difference between the bias correction network and the parallel physics network.
Due to the diverse complexity of different ecosystems, the underlying processes represented in the data used for on-site temporal prediction and the data used for multi-site generalising prediction differ. 
In the latter objective, PGNNs have a yet increased advantage over the PM and the naive neural network (see \ref{fig:performance_temporal}, c and d).
In the PM, correlations of the exogenous variables are fixed by its twenty parameters which are biased through missing processes. In a multi-site calibration, the PM has to capture the site-specific correlations, increasing this bias even more.
Equivalently, the PGNN is forced to find similarities and fit cross-site correlations in multi-site prediction. This increases the constraint that originates from the data, demanding stronger causality in the represented relationships. 
In contrast to the PGNN, the PM thus performs constrained by its parameters as well as site-specific data quality and fit of the theory.

\subsection*{Detecting process mis-specification}

One drawback in the application of deep neural networks and as such also of integrated approaches is the lacking interpretability of the learned model parameters and the causes for a specific prediction. The field of explainable machine learning is just emerging, yet simple model-agnostic tools applicable for interpreting regression models can also be applied to neural networks and PGNNs \citep{molnar2022}. 
With correlated input variables, individual conditional expectation (ICE) plots allow quick assessment of variable importance and uncover potential interactions among them.
The parallel physics network is a particularly interesting example for this interpretability analysis. By design it is forced to fit the residuals of the PM (see above). 
The ICE plots show the effect of each variable on the variance in the data that remained unexplained by the PM (see Fig.~\ref{fig:GPPeffectsize}). Both, the naive neural network and the parallel physics network analysis suggest a strong linear effect of the vapour pressure deficit $D$, while this effect is nearly undetectable in the PM. In the structure of the PM, the light-use efficiency is the driving relationship and $D$ is considered only through one of multiple modifiers that adjusts it. The modifier for $D$ now is mainly dependent on one parameter ($\kappa$). 
Of course we cannot causally draw the conclusion that the PM is mis-specified with regard to $D$, the processes that the PGNN is fitting remain unknown.
Yet it records a trace that can be complemented with the results from the Bayesian calibration of the PM. A wide posterior distribution of $\kappa$ and its ambiguous convergence points towards a non-identifiability of this parameter (see Supplementary Information).
In addition to a variable-based interpretability analysis, we propose to examine prediction within fixed or moving periods in which specific intra-annual variations in the modeled process can be expected \citep{wu2012effects}. 
Such post-hoc analyses can deliver insights into processes that are represented in a PGNN but missing in a pure PM. Findings can then be experimentally investigated and eventually incorporated back into the PM. We thus suggest the use of PGNNs not only to deliver improved predictive performance, but for aiming further: to detect wholes in the theories we work with and to extend their foundations.

\subsection*{Technical hurdles and successes}

Building a PGNN from a PM and a neural network involves manifold technical decisions in both domains.
The key decisions for an analysis with neural networks are on the choice of the optimisation algorithm, its hyper-parameters and the model architecture. Further, on the model type itself and on the splitting of the data. Part of this are already overcome with the neural architecture search, but additional manual finetuning is often required. Suboptimal decisions are reflected by untrainable folds in the cross-validation or in exessive rashes of the validation error during the training procedure.
To simplify further the remaining choices, we propose to automatise the decision process using methods grouped under the field of automated machine learning (Auto-ML), for which user-friendly applications already exist \citep{mendozaautomlbook18a}. 
The PM in this case study is with its just under thirty parameters comparatively simple regarding computational effort and data requirements. 
This is not the standard in ecological and biogeochemical modelling, which may limit the practical application of some of the PGNN approaches, i.e. the embedded network.
For more complex PMs that do not lend themselves to repeated calibration, we recommend using one of the PGNN approaches based only on PM predictions.
The complexity of the PM is likewise relevant in generating synthetic data, which is used for pre-training in the domain adaptation. For sampling, either informative prior distributions can be used that are narrow around a default calibration, or alternatively, uninformative priors can be used that are flat within a pre-defined parameter range. 
This decision affects strength of the constraint imposed onto the neural network, as well as the amount of data that must be used for pre-training.

\subsection*{Conclusion}

The comparison of process-based neural networks shows a categorised sample of existing methods from an integrated modelling field that we find especially useful for environmental modelling, regarding the data-theory spectrum. This area of research is developing fast in the area of fundamental physical applications and in working with differential equations \citep{karniadakis2021physics}. We have presented here the specific relevance of PGNNs for classical prediction tasks in the environmental sciences. Furthermore, we justify integrated approaches from two perspectives: Embedding the processes in a neural network may allow for their better interpretability, while simultaneously addressing mis-specification in the process models. More complex process models will certainly demand refinement of the demonstrated methodology.





\acknow{This work profitted from discussion with Nuno Carvalhais.} 

\showacknow 

\bibliography{PMinNN}

\begin{thebibliography}{10}

\bibitem{Asner2020}
GP Asner, et~al., Large-scale mapping of live corals to guide reef
  conservation.
\newblock {\em\protect\JournalTitle{Proceedings of the National Academy of
  Sciences}} \textbf{117}, 33711–33718 (2020).

\bibitem{Brandt2020}
M Brandt, et~al., An unexpectedly large count of trees in the {W}est {A}frican
  {S}ahara and {S}ahel.
\newblock {\em\protect\JournalTitle{Nature}} \textbf{587}, 78–82 (2020).

\bibitem{Duporge2021}
I Duporge, O Isupova, S Reece, DW Macdonald, T Wang, Using very-high-resolution
  satellite imagery and deep learning to detect and count {African} elephants
  in heterogeneous landscapes.
\newblock {\em\protect\JournalTitle{Remote Sensing in Ecology and
  Conservation}} \textbf{in press} (2021).

\bibitem{Jarriel2021}
T Jarriel, J Swartz, P Passalacqua, Global rates and patterns of channel
  migration in river deltas.
\newblock {\em\protect\JournalTitle{Proceedings of the National Academy of
  Sciences}} \textbf{118} (2021).

\bibitem{Bury2021}
TM Bury, et~al., Deep learning for early warning signals of tipping points.
\newblock {\em\protect\JournalTitle{Proceedings of the National Academy of
  Sciences}} \textbf{118} (2021).

\bibitem{Reichstein2019}
M Reichstein, et~al., Deep learning and process understanding for data-driven
  {E}arth system science.
\newblock {\em\protect\JournalTitle{Nature}} \textbf{566}, 195–204 (2019).

\bibitem{Dormann2012}
CF Dormann, et~al., Correlation and process in species distribution models:
  bridging a dichotomy.
\newblock {\em\protect\JournalTitle{Journal of Biogeography}} \textbf{39},
  2119–2131 (2012).

\bibitem{Bongard2007}
J Bongard, H Lipson, Automated reverse engineering of nonlinear dynamical
  systems.
\newblock {\em\protect\JournalTitle{Proceedings of the National Academy of
  Sciences of the United States of America}} \textbf{104}, 9943–9948 (2007).

\bibitem{Schmidt2013}
M Schmidt, H Lipson, Distilling free-form natural laws.
\newblock {\em\protect\JournalTitle{Science}} \textbf{81}, 81–85 (2013).

\bibitem{Wood2001}
SN Wood, Partially specified ecological models.
\newblock {\em\protect\JournalTitle{Ecological Monographs}} \textbf{71}, 1–25
  (2001).

\bibitem{Rackauckas2021}
C Rackauckas, R Sharma, B Lie, Hybrid mechanistic + neural model of laboratory
  helicopter in {\em Proceedings of The 61st SIMS Conference on Simulation and
  Modelling SIMS 2020, September 22-24, Virtual Conference, Finland}.
\newblock (Linköping Electronic Conference Proceedings), Vol.{} 176, p.
  264–271 (2021).

\bibitem{Clark2006}
JS Clark, AE Gelfand, A future for models and data in environmental science.
\newblock {\em\protect\JournalTitle{Trends in Ecology and Evolution}}
  \textbf{21}, 375–80 (2006).

\bibitem{Hartig2012}
F Hartig, et~al., Connecting dynamic vegetation models to data - an inverse
  perspective.
\newblock {\em\protect\JournalTitle{Journal of Biogeography}} \textbf{39},
  2240–2252 (2012).

\bibitem{Dietze2017}
MC Dietze, {\em Ecological Forecasting}.
\newblock (Princeton University Press, Princeton, N.J.), (2017).

\bibitem{Fer2018}
I Fer, et~al., Linking big models to big data: efficient ecosystem model
  calibration through bayesian model emulation.
\newblock {\em\protect\JournalTitle{Biogeosciences}} \textbf{15}, 5801–5830
  (2018).

\bibitem{Speich2021}
M Speich, CF Dormann, F Hartig, Sequential {M}onte-{C}arlo algorithms for
  {B}ayesian model calibration – a review and method comparison.
\newblock {\em\protect\JournalTitle{Ecological Modelling}} \textbf{455}, 109608
  (2021).

\bibitem{Ba2019}
Y Ba, G Zhao, A Kadambi, Blending diverse physical priors with neural networks.
\newblock {\em\protect\JournalTitle{arXiv}} \textbf{[cs, stat]}, 1910.00201
  (2019).

\bibitem{Karpatne2018}
A Karpatne, W Watkins, J Read, V Kumar, Physics-{G}uided {N}eural {N}etworks
  ({PGNN}): An application in lake temperature modeling.
\newblock {\em\protect\JournalTitle{arXiv}} \textbf{[physics, stat]},
  1710.11431v2 (2018).

\bibitem{Zhao2019}
WL Zhao, et~al., Physics-constrained machine learning of evapotranspiration.
\newblock {\em\protect\JournalTitle{Geophysical Research Letters}} \textbf{46},
  14496–14507 (2019).

\bibitem{karniadakis2021physics}
GE Karniadakis, et~al., Physics-informed machine learning.
\newblock {\em\protect\JournalTitle{Nature Reviews Physics}} \textbf{3},
  422--440 (2021).

\bibitem{reyer2020profound}
CP Reyer, et~al., The profound database for evaluating vegetation models and
  simulating climate impacts on european forests.
\newblock {\em\protect\JournalTitle{Earth System Science Data}} \textbf{12},
  1295--1320 (2020).

\bibitem{peltoniemi2015semi}
M Peltoniemi, et~al., {\em A semi-empirical model of boreal-forest gross
  primary production, evapotranspiration, and soil water-calibration and
  sensitivity analysis}.
\newblock (Finnish Environment Institute), (2015).

\bibitem{makela2008developing}
A M{\"a}kel{\"a}, et~al., Developing an empirical model of stand gpp with the
  lue approach: analysis of eddy covariance data at five contrasting conifer
  sites in europe.
\newblock {\em\protect\JournalTitle{Global Change Biology}} \textbf{14},
  92--108 (2008).

\bibitem{Gongora2019}
JA Gongora, HP Marshall, C Hase, T Bradberry, S Ballerini, Applications of deep
  learning and domain adaptation to the study of mountain snow.
\newblock {\em\protect\JournalTitle{AGU Fall Meeting Abstracts}} \textbf{33}
  (2019).

\bibitem{roberts2017cross}
DR Roberts, et~al., Cross-validation strategies for data with temporal,
  spatial, hierarchical, or phylogenetic structure.
\newblock {\em\protect\JournalTitle{Ecography}} \textbf{40}, 913--929 (2017).

\bibitem{hartig2019bayestools}
F Hartig, F Minunno, S { Paul}, {\em BayesianTools: General-Purpose MCMC and
  SMC Samplers and Tools for Bayesian Statistics}, (2019) R package version
  0.1.7.

\bibitem{terBraak2008}
CJF ter Braak, JA Vrugt, {Differential Evolution Markov Chain} with snooker
  updater and fewer chains.
\newblock {\em\protect\JournalTitle{Statistics and Computing}} \textbf{18},
  435–446 (2008).

\bibitem{molnar2022}
C Molnar, {\em Interpretable Machine Learning}.
\newblock 2 edition, (2022).

\bibitem{wu2012effects}
J Wu, et~al., Effects of climate variability and functional changes on the
  interannual variation of the carbon balance in a temperate deciduous forest.
\newblock {\em\protect\JournalTitle{Biogeosciences}} \textbf{9}, 13--28 (2012).

\bibitem{mendozaautomlbook18a}
H Mendoza, et~al., Towards automatically-tuned deep neural networks in {\em
  AutoML: Methods, Sytems, Challenges}, eds.{} F Hutter, L Kotthoff, J
  Vanschoren.
\newblock (Springer), pp. 141--156 (2018).

\end{thebibliography}


\begin{thebibliography}{}

\bibitem[Minunno et~al., 2016]{minunno2016preles}
Minunno, F., Peltoniemi, M., Launiainen, S., Aurela, M., Lindroth, A., Lohila,
  A., Mammarella, I., Minkkinen, K., \& M{\"a}kel{\"a}, A. (2016).
\newblock Calibration and validation of a semi-empirical flux ecosystem model
  for coniferous forests in the boreal region.
\newblock {\em Ecological Modelling}, 341, 37--52.

\bibitem[Reyer et~al., 2019]{Reyer2019}
Reyer, C., Gonzalez, R.~S., Dolos, K., Hartig, F., Hauf, Y., Noack, M.,
  Lasch-Born, P., Rötzer, T., Pretzsch, H., Meesenburg, H., et~al. (2019).
\newblock The {PROFOUND} database for evaluating vegetation models and
  simulating climate impacts on forests.
\newblock {\em GFZ Data Services}, 0.1.12.

\bibitem[Taiz \& Zeiger, 2015]{Taiz2015}
Taiz, L. \& Zeiger, E. (2015).
\newblock {\em Plant Physiology}, chapter Photosynthesis: Physiological and
  Ecological Considerations, (pp.\ 172--174).
\newblock Sinauer Associates, Inc., Publishers: Sunderland, Massachusetts, 6
  edition.

\end{thebibliography}

\end{document}


\maketitle
\renewcommand{\contentsname}{Table of Contents}
\makeatletter
\@starttoc{toc}
\makeatother

\newpage
\section{Pipeline}
\subsection{Structure}
\begin{figure}[htb]
\centering
\includegraphics[scale=0.8]{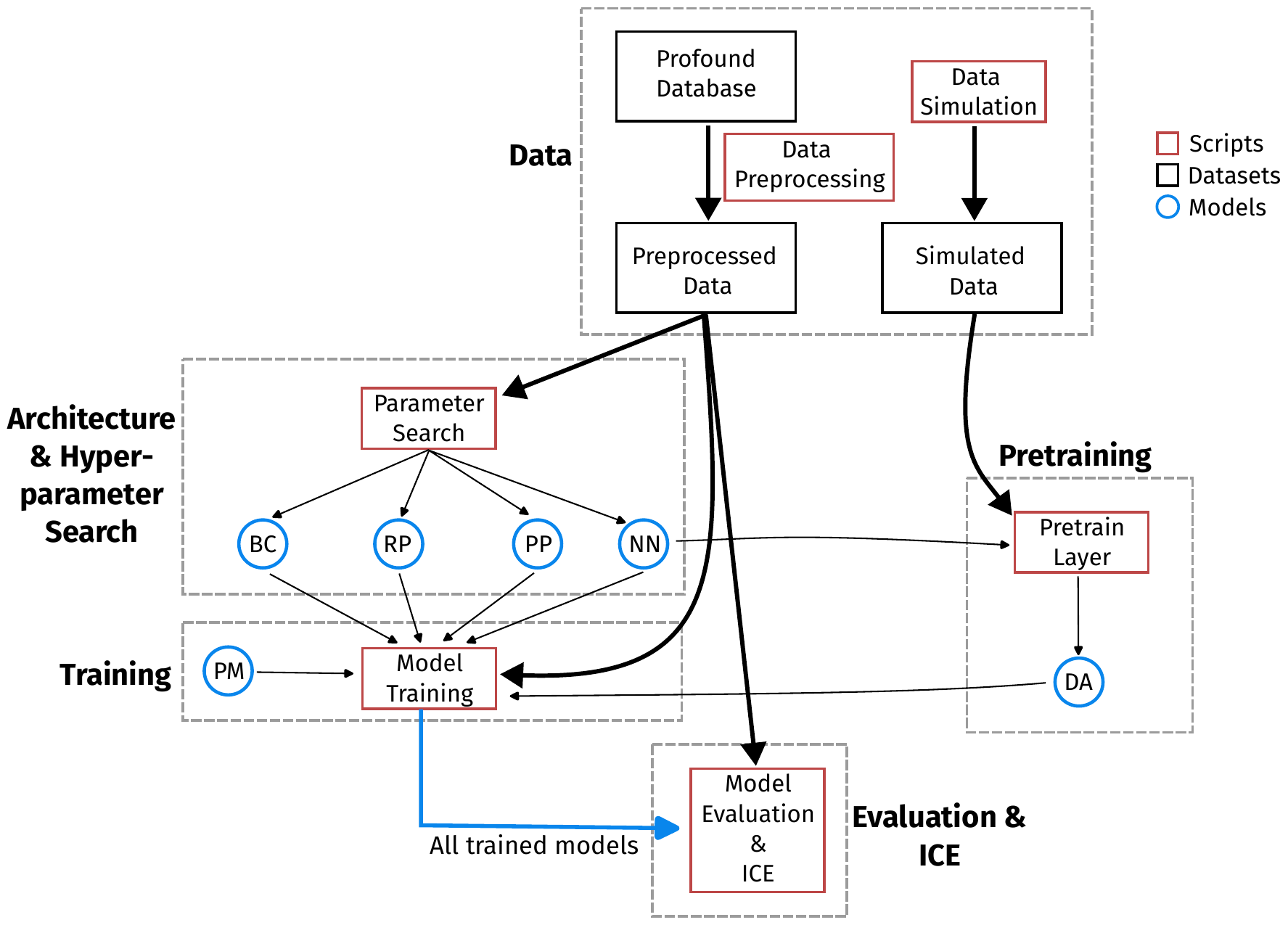}
\caption{Sketch of the pipeline structure. The sequential order of the parts is from top to bottom. The structure is similar for the temporal and spatial experiment. PM = Process model, BC = Bias correction, RP= Residual physics, PP = Parallel physics, NN = Naive neural network, DA = Domain adaptation}
\label{fig:struc}
\end{figure}

The pipeline structure is sketched in Fig. \ref{fig:struc}.

\subsection{Download code}
The code pipeline can be downloaded from GitHub \url{https://github.com/therealniklasmoser/physics_guided_nn}.


The pipeline consists of the folders src, data, code and r. The modified source code of PRELES is stored in the src folder. To compile the modified PRELES code to a PyTorch cpp-extension, that is callable from Python the source code needs to be compiled. 
We use the compiler version that is shown below.
\begin{lstlisting}
 niklas@hpc:~$ gcc --version
 gcc (GCC) 9.1.0
 Copyright (C) 2019 Free Software Foundation, Inc.
 This is free software; see the source for copying conditions.  There is NO
 warranty; not even for MERCHANTABILITY or FITNESS FOR A PARTICULAR PURPOSE. 
  
\end{lstlisting}

To convert the modified PRELES c-code to a proper PyTorch cpp-extension run the setup.py script in the src folder using the command:

\begin{lstlisting}
niklas@hpc:~$ python setup.py install
running install
running bdist_egg
running egg_info
writing preles.egg-info/PKG-INFO
writing dependency_links to preles.egg-info/dependency_links.txt
writing top-level names to preles.egg-info/top_level.txt
reading manifest file 'preles.egg-info/SOURCES.txt'
writing manifest file 'preles.egg-info/SOURCES.txt'
installing library code to build/bdist.linux-x86_64/egg
running install_lib
running build_ext
building 'preles' extension
...
Processing dependencies for preles==0.0.0
Finished processing dependencies for preles==0.0.0
\end{lstlisting}

After running the setup script one can test if PRELES is callable in Python by the commands below. To call PRELES it is necessary to import torch first.

\begin{lstlisting}
niklas@hpc:~$ python
Python 3.7.10 | packaged by conda-forge | (default, Feb 19 2021, 16:07:37)
[GCC 9.3.0] on linux
Type "help", "copyright", "credits" or "license" for more information.

import torch
import preles

\end{lstlisting}

\section{Data}
\subsection{Source}
We use the PROFOUND database. It provides information
about the vegetation and climate at the forest stand scale in time series consisting of data of several years for forest sites in Europe\citet{Reyer2019}. The PROFOUND database combines empirical data about forest stand
structure, species composition, climatic conditions, energy and mineral fluxes. The data can be downloaded from the GitHub \url{https://github.com/COST-FP1304-PROFOUND/ProfoundData}.

\subsection{Preprocessing}
For the analysis, the same input variables are used in the same units as in PRELES. The variables needed are namely the daily sums of photosynthetic active radiation (PAR), the mean air temperature (TAir), the vapour pressure deficit (VPD), the precipitation above canopy (Precip), the carbon dioxide concentration of air (CO$_2$), the fraction of photosynthetic active radiation (fAPAR) and the day of year (DOY), as depicted in Table~\ref{tab: vars}. Additionally, information on the gross primary production (GPP) is required. It is used as the output variable in the models. The data required is in a daily resolution.

\begin{table}[htb]
	\caption{Data used from the PROFOUND database.} \label{tab: vars}	
	\centering
	\begin{tabular}{lcc} 
		\toprule
		Abbreviation  & Description  & Unit \\ 
		\midrule
		PAR   & sum of photosynthetic active radiation & $\si{mol.m^{-2}.d^{-1}}$ \\
		TAir & mean air temperature & $\si{\degree C}$ \\
		VPD & mean vapour pressure deficit & $\si{kPa}$  \\ 
		Precip & precipitation above canopy & $\si{mm}$  \\
		CO$_2$ & carbon dioxide concentration & $\si{ppm}$ \\
		fAPAR & fraction of photosynthetic active radiation & - \\
		GPP & gross primary production & $\si{g.C.m^{-2}.d^{-1}}$ \\ 
		\bottomrule
	\end{tabular}
\end{table}

The PROFOUND database information on the irradiance, the global radiation $\phi_\text{e}$ (in $\si{J.cm^{-2}.d^{-1}}$) is converted into the quantum units describing the photon irradiance $\phi_\text{p}$ (in $\si{mol.m^{-2}.d^{-1}}$). Therefore, the relation of the wavelength and the energy of a photon is used as described in \citep{Taiz2015} with

\begin{equation}
	\phi_\text{p} = \frac{\phi_\text{e}}{E N_\text{A}} \quad \text{,}
\end{equation}
and
\begin{equation}
E = \frac{h c}{\lambda} \quad \text{,}
\end{equation} 

where $h$ is Planck's constant ($6.63 \cdot 10^{-34} \si{J.s}$), $c$ is the speed of light ($2.99792458 \cdot 10^8 \si{m.s^{-1}}$), $\lambda$ is the wavelength ($\approx 2.2 \cdot 10^{-7} \si{m}$), $N_\text{A}$ is Avogadro's constant ($6.602 \cdot 10^{23} \si{mol^{-1}}$) and $\phi_\text{e}$ is the global radiation (in $\si{J.s^{-1}.m^{-2}}$).

fAPAR is derived from MODIS satellite data and has a resolution of eight days. The data gaps of the eight day resolution data are filled by averaging information of the data point before and after the gap. Second, to get the daily resolution, it is assumed that the eight day value is representative for all days within this period. Therefore, fAPAR is set constant over the eight days.

GPP is converted from $\si{\mu mol.CO_{2}.m^{-2}.s^{-1}}$ to $\si{g.C.m^{-2}.d^{-1}}$ using the molar mass of carbon ($\approx 12.011 \si{g.mol^{-1}}$).




\subsection{Normalisation}
Each variable is normalised around its mean $\sigma$ and its standard deviation $\sigma$ before using it as an input in the neural networks. Therefore, a $z$-score is calculated for each variable as
\begin{equation}
	z = \frac{x - \mu}{\sigma} \quad \text{.}
\end{equation}

Because of its cyclic character a sine and cosine function is used to normalise DOY ($d$). This ensures that the information of the last day of  year $n$ and the first day of year $n+1$ are closer together (compared to $365$ and $1$). DOY is transformed as
\begin{equation}
	d_\text{sin} = \sin{\left(d \frac{2\pi}{365}\right)} \quad \text{,}
\end{equation}
and
\begin{equation}
d_\text{cos} = \cos{\left(d \frac{2\pi}{365}\right)} \quad \text{.}
\end{equation}

\section{Architecture and hyperparameter search}
We implemented a combined architecture and hyper-parameter search. The architecture search space consists of $300$ randomly sampled layersizes with a maximum depth of $4$ hidden layers. The hyper-parameter search space consists of $300$ randomly sampled hyper-parameter vectors. In the combined search each network of the $300$ architecture candidates is fitted using the $300$ candidate hyper-parameter vectors. Each hyper-parameter vector consists of learning rate, batch size. For the physics embedding and the physics regularisation the hyper-parameter vectors consist additionally of a regularisation factor. The best performing candidate network $k$ is chosen based on the minimum of the index $i_k$ with
\begin{equation}
i_k = \frac{\mathbf{E}\left[\mathcal{L}_{\text{val}}\right]^2 + \sqrt{\mathbf{E}\left[(\mathcal{L}_{\text{val}}- \mathbf{E}\left[\mathcal{L}_{\text{val}}\right])^2\right]}}{2} \quad \text{,} 
\end{equation}
where $\mathcal{L}_{\text{val}}$ denotes the validation losses for all cross-validation runs.

The best performing architecture and hyper-parameters are shown for each model in Tab.  \ref{tab.:NAS} for the full data experiment and in Tab. \ref{tab: varss} for the sparse data experiment.

\begin{table}
	\caption{Full data architecture and hyper-parameters for each model and the temporal experiment} \label{tab: vars}	
	\centering
	\begin{tabular}{lccccc} 
		\toprule
		Model  & Architecture & LR & BS & $\lambda$ & $i$ \\ 
		\midrule
		NN  & $[2, 64, 128]$ & $0.0143$ & $16$ & - & $0.306126$ \\
		BC & $[8, 64]$ & $0.0531$ & $16$ & - & $2.928012$ \\
        PP &  $[256, 128]$ & $0.002$ & $16$ & - & $1.769507$ \\
        RP &  $[256, 32, 8]$ & $0.0265$ & $4$ & $0.0101$ & $0.584885$ \\
		\bottomrule
	\end{tabular}
 \label{tab.:NAS}
\end{table}

\begin{table}
	\caption{Sparse data architecture and hyper-parameters for each model and the temporal experiment} \label{tab: varss}	
	\centering
	\begin{tabular}{lccccc} 
		\toprule
		Model  & Architecture & LR & BS & $\lambda$ & $i$ \\ 
		\midrule
		NN  & $[2, 8, 64]$ & $0.0082$ & $2$ & - & $0.943277$ \\
		BC & $[16, 256]$ & $0.0122$ & $4$ & - & $1.371335$ \\
        PP &  $[2]$ & $0.0735$ & $64$ & - & $3.362334$ \\
        RP &  $[2, 256,8, 256]$ & $0.0184$ & $16$ & $0.0654$ & $1.354255$ \\
		\bottomrule
	\end{tabular}
 \label{tab.:NAS}
\end{table}

\begin{figure}
\centering
\includegraphics[scale=0.5]{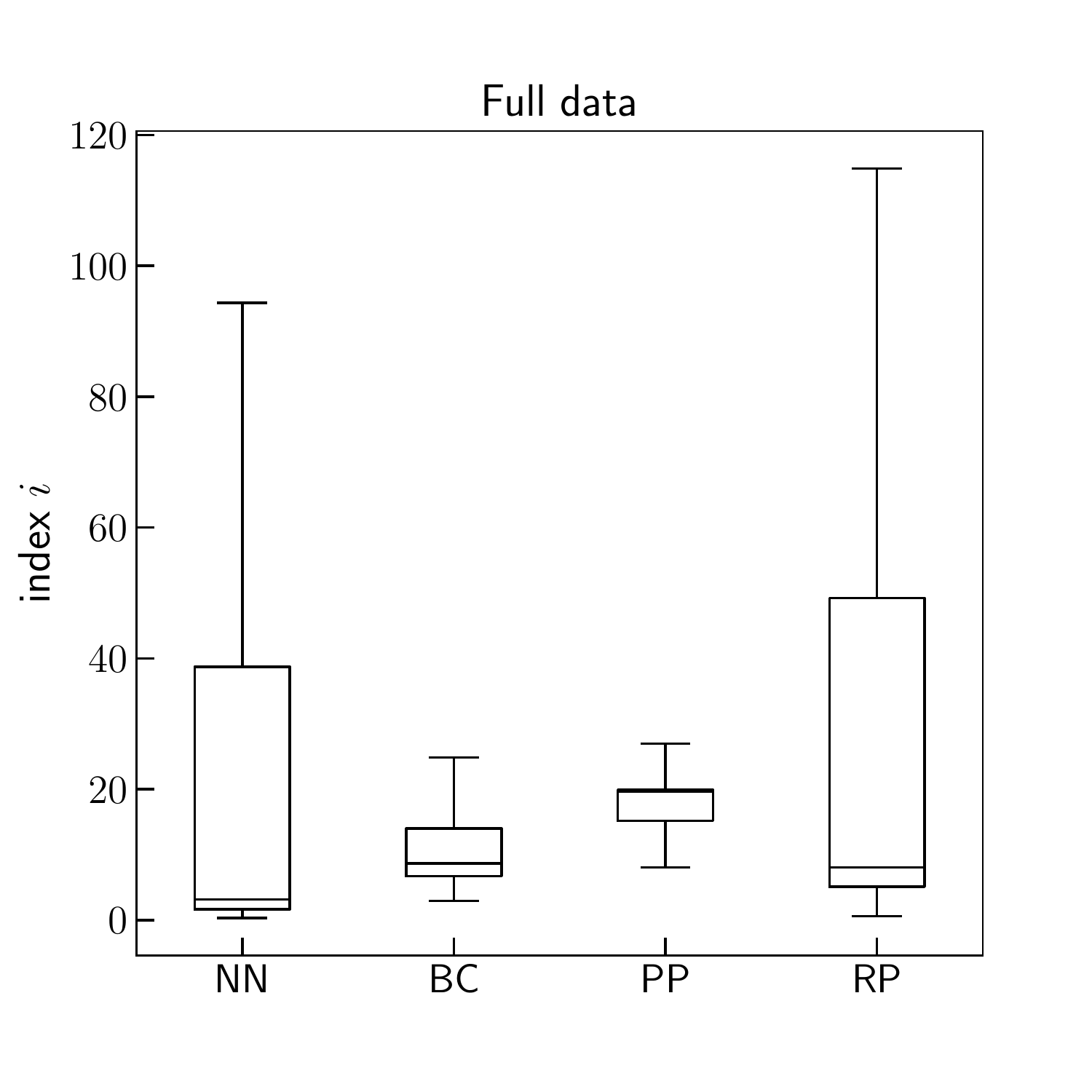}
\includegraphics[scale=0.5]{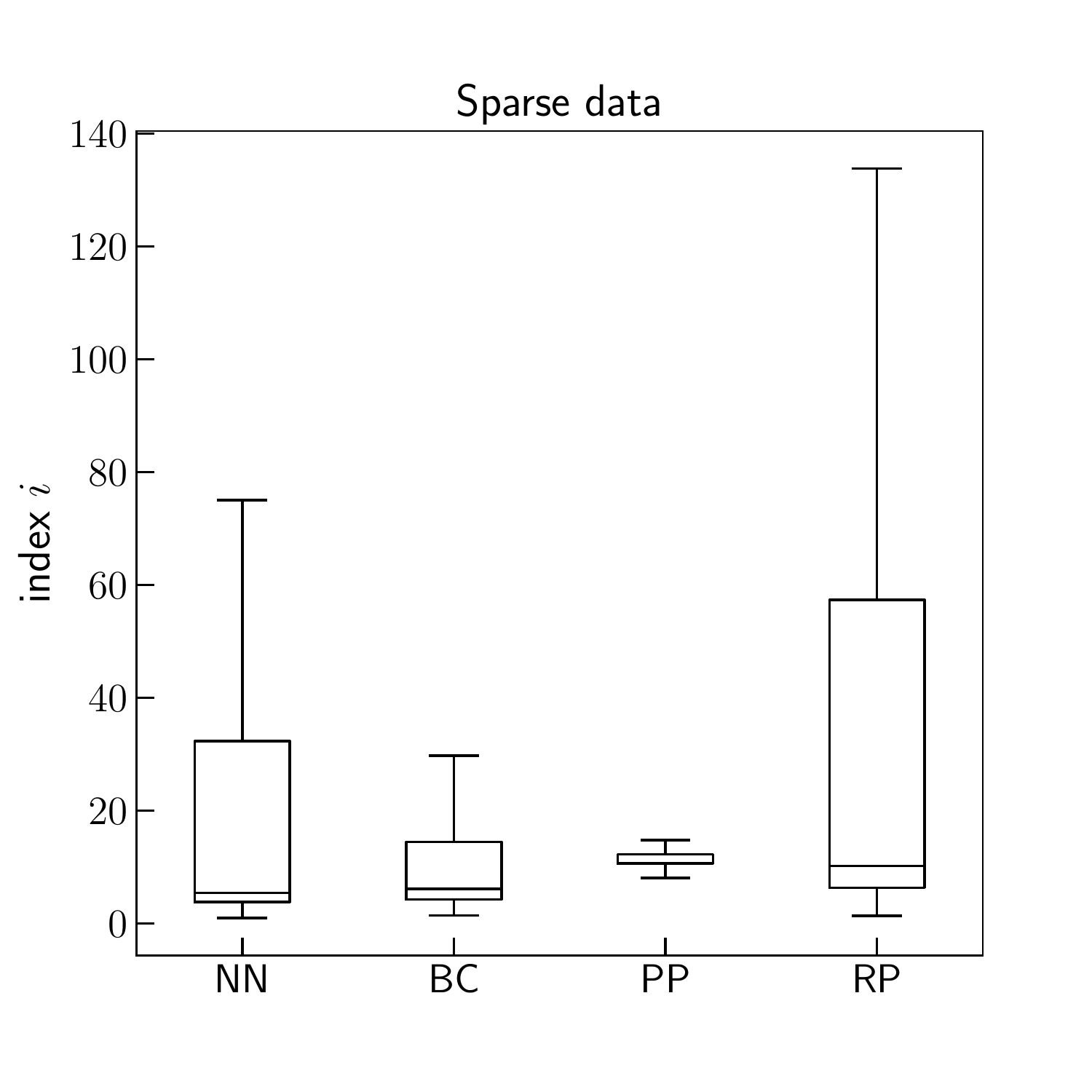}
\caption{Index values for all candidate models in the architecutre and hyper-parameter search. PM = Process model, BC = Bias correction, RP= Residual physics, PP = Parallel physics, NN = Naive neural network, DA = Domain adaptation}
\label{fig:bp}
\end{figure}

The performance of the candidate architectures and hyper-parameters are shown in Fig. \ref{fig:bp} for the temporal experiment. 

\section{Technical details of the domain adaptation}

The yearly equivalents of the neural network training data were the basis for the climate simulations. Except for CO$_2$, which was fixed at a value of 380 ppm, the five climatic variables $y = \{\text{T, $\phi$, D, P, $f_\text{aPPFD}$}\}$ were separately modeled with a generalized additive model (GAM).
For the on-site simulations, each $y_k \in y$ was described as a function of the day of the year (DOY) in interaction with the year $R$. The GAM was fitted with a cyclic cubic smooth function $f$ imposed onto DOY. Allowing for the interaction with $R$, the smooth varies with the year.
\begin{equation}
    y_k = \beta_{0} + \sum_{i=1}^{4} f_{i}(\text{DOY})\cdot R_{i} + \epsilon
\end{equation}
For the multi-site simulations, each $y_k$ was described as a function of the day of the year (DOY) in interaction with the site $S$.
\begin{equation}
    y_k = \beta_{0} + \sum_{i=1}^{5} f_{i}(\text{DOY})\cdot S_{i} + \epsilon
\end{equation}
The resulting models could be used to simulate estimated time series of the variables of any length and for any time point of years and sites under consideration.
To introduce variation to the simulation, noise $\nu$ was added to each $\hat{y}_k$, sampled from a multivariate normal distribution $\nu \sim \mathcal{N}_{k=5}(\mu, \Sigma)$. Here, $\mu$ denotes the $k$-dimensional mean vector and $\Sigma$ is the $k\times k$-dimensional covariance matrix. Means $\mu$ were set to zeros while the covariance matrix $\Sigma$ was specified explicitly as the covariance matrix of the $k$-dimensional residual matrix.

With both, a global and local sensitivity analyses, the five most sensititive stand specific PRELES parameters were considered. These are the potential light use efficiency ($\beta$), the threshold for the state of acclimation ($X0$), the light modifier parameter for saturation with irradiance ($\gamma$), and the transpiration and evaporation parameters ($\alpha$ and $\chi$).
Prior knowledge about their marginal distributions is available as uniform prior distribution parameters, used for the Bayesian calibration of PRELES \citep{minunno2016preles}. In order to provide the neural network with the same information as PRELES, the prior distributions were strongly narrowed down, ressembling the default calibration.
In a Latin hypercube design the five parameters were sampled from their neat uniform joint probability distribution. The other parameters remained fix at their default values.

\begin{table}
\centering
\caption{R$^2$ statistics of fitted GAMs for synthetic data generation under sparse data assumption.}
\begin{tabular}{lrr}
Variable & On-site R$^2$ & Multi-site R$^2$  \\
\midrule
$T$ &0.9836  & 0.7427 \\
$R$ & 0.9791 & 0.3995 \\
$D$ & 0.9582 &  0.7145 \\
$\phi$ & 0.9928 &  0.7888 \\
$f_\text{aPPFD}$ & 0.9771 & 0.6761 \\
\bottomrule
\end{tabular}
\label{tab:gamstats}
\end{table}

\section{Technical details of the physics embedding}
We convert the PRELES c-code into a callable Python library. Therefore, we change the elementary operations in the c-code to their PyTorch equivalent, e.g. \lstinline!sin(x)!$\rightarrow$\lstinline!torch.sin(x)!. This is a necessary condition to use PRELES as a forward pass of the neural network. In the PyTorch framework all calculations on the input tensor are stored in a computational graph during forward propagation. During backpropagation the computational graph of the tensor is used to numerically calculate intermediate gradients for all elementary operations on the tensor.

\section{Temporal predictions}
\begin{figure}[htb]
\centering
\includegraphics[scale=0.42]{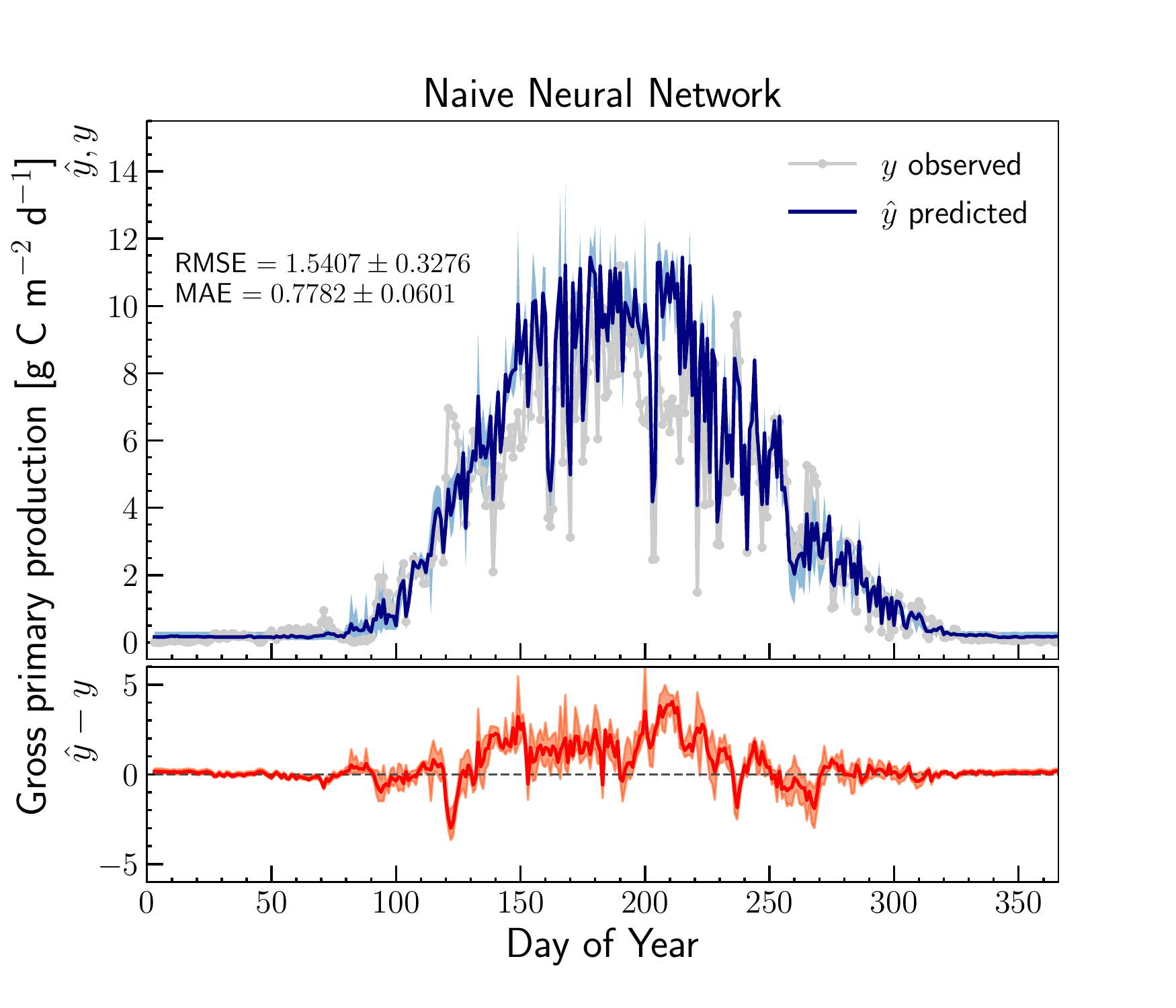}
\includegraphics[scale=0.42]{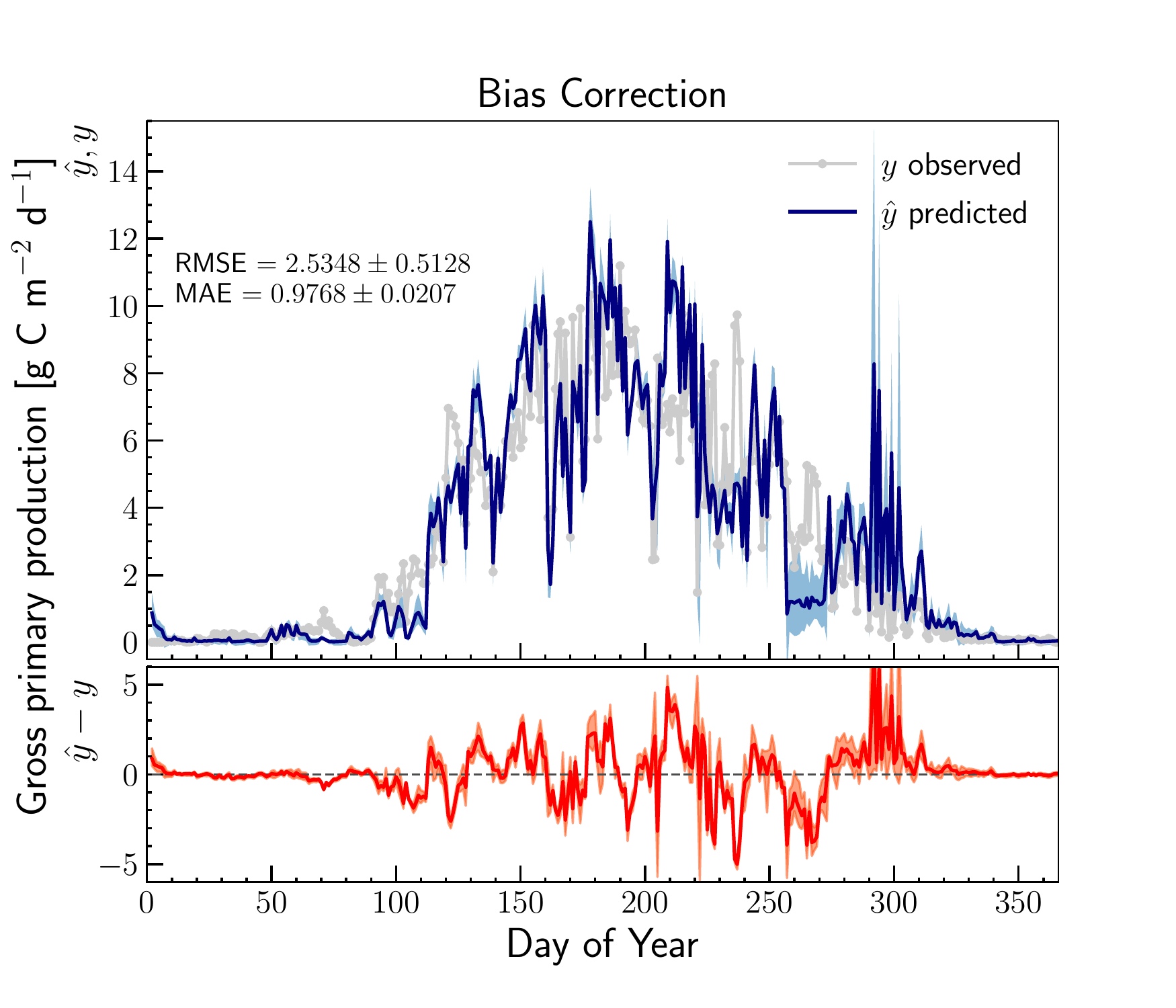}
\includegraphics[scale=0.42]{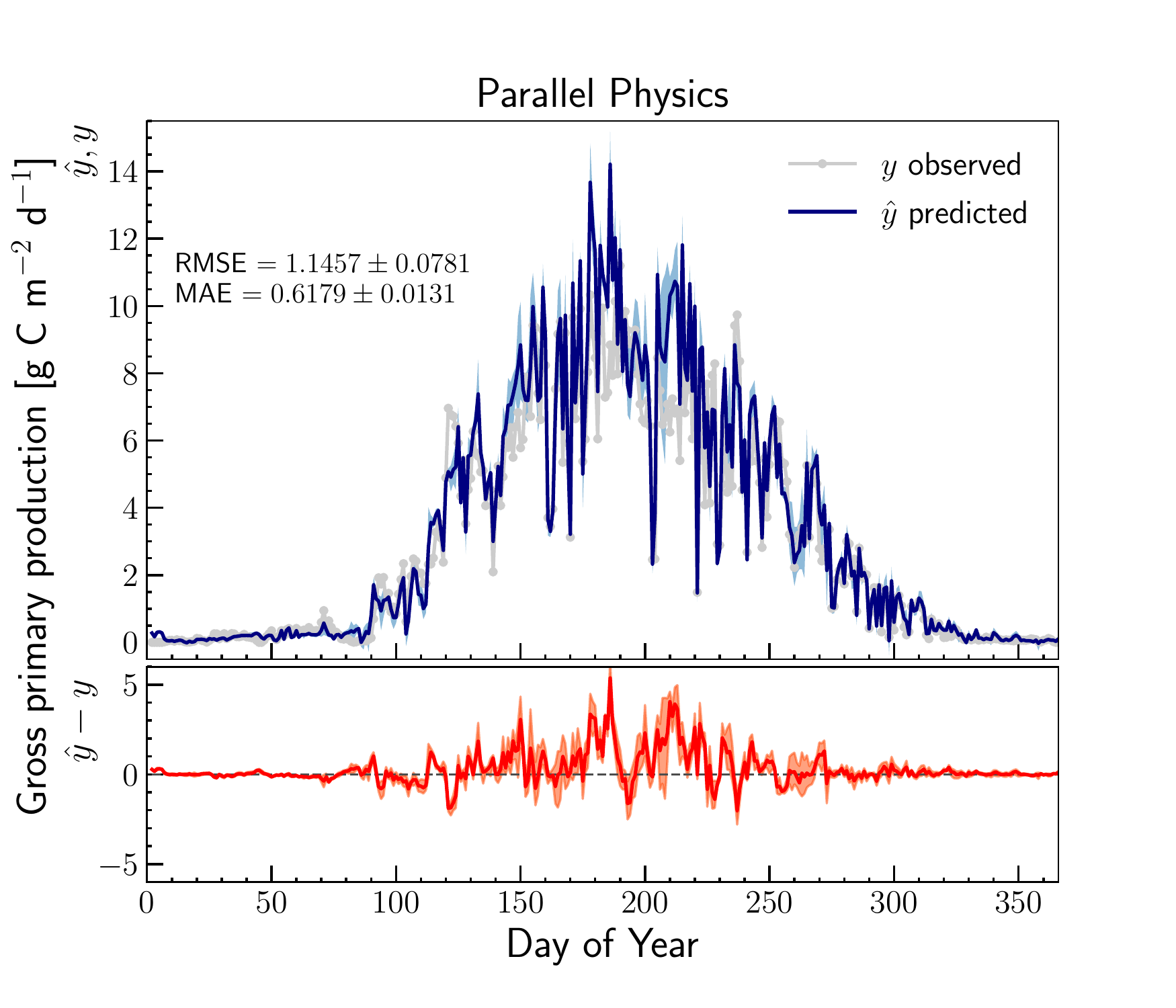}
\includegraphics[scale=0.42]{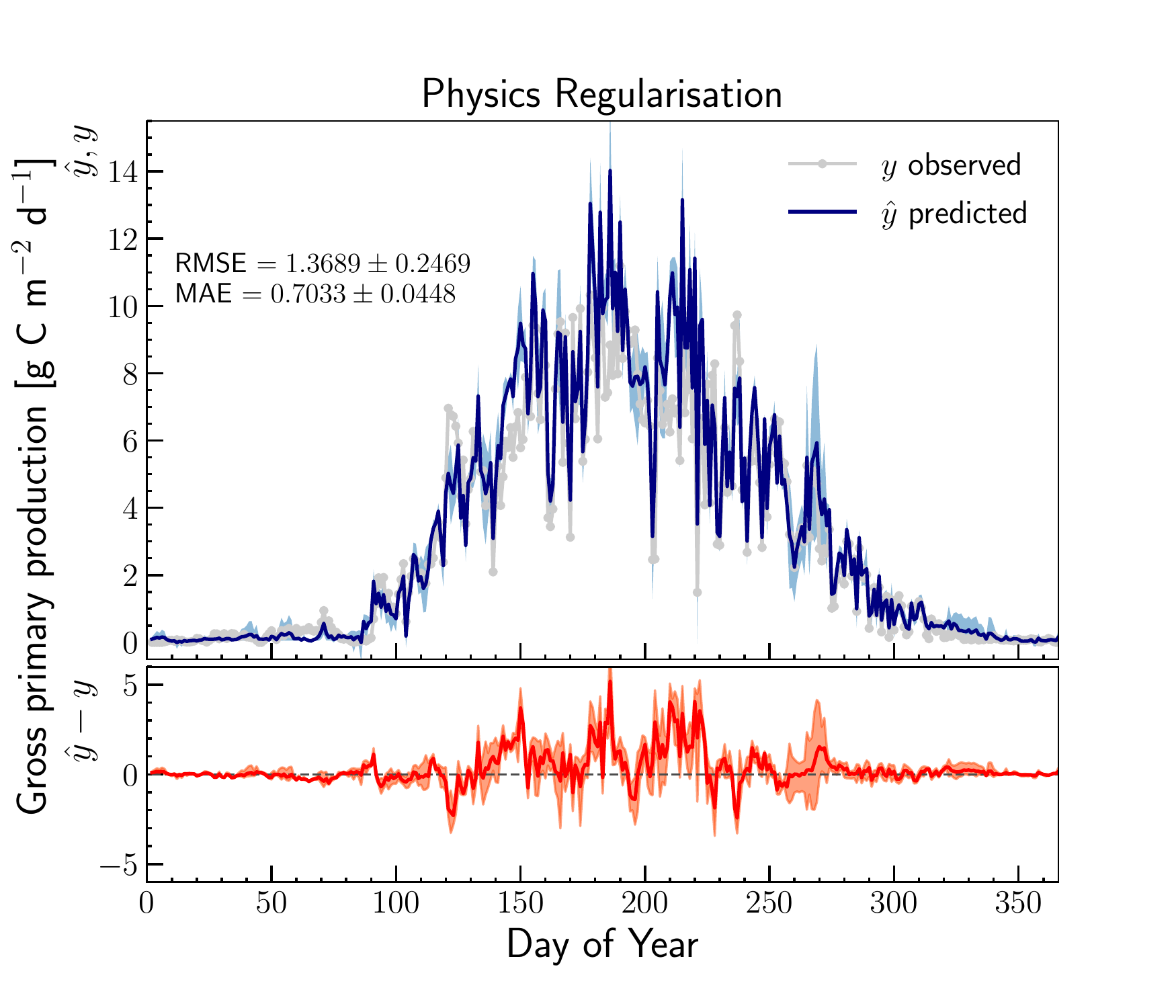}
\caption{Temporal predictions with the full data set.}
\label{fig:temp}
\end{figure}

The temporal predictions with the full data set are shown in Fig. \ref{fig:temp}. 

\section{Bayesian Calibration of the process model}

\begin{figure}
    \centering
    \includegraphics[width=.99\textwidth]{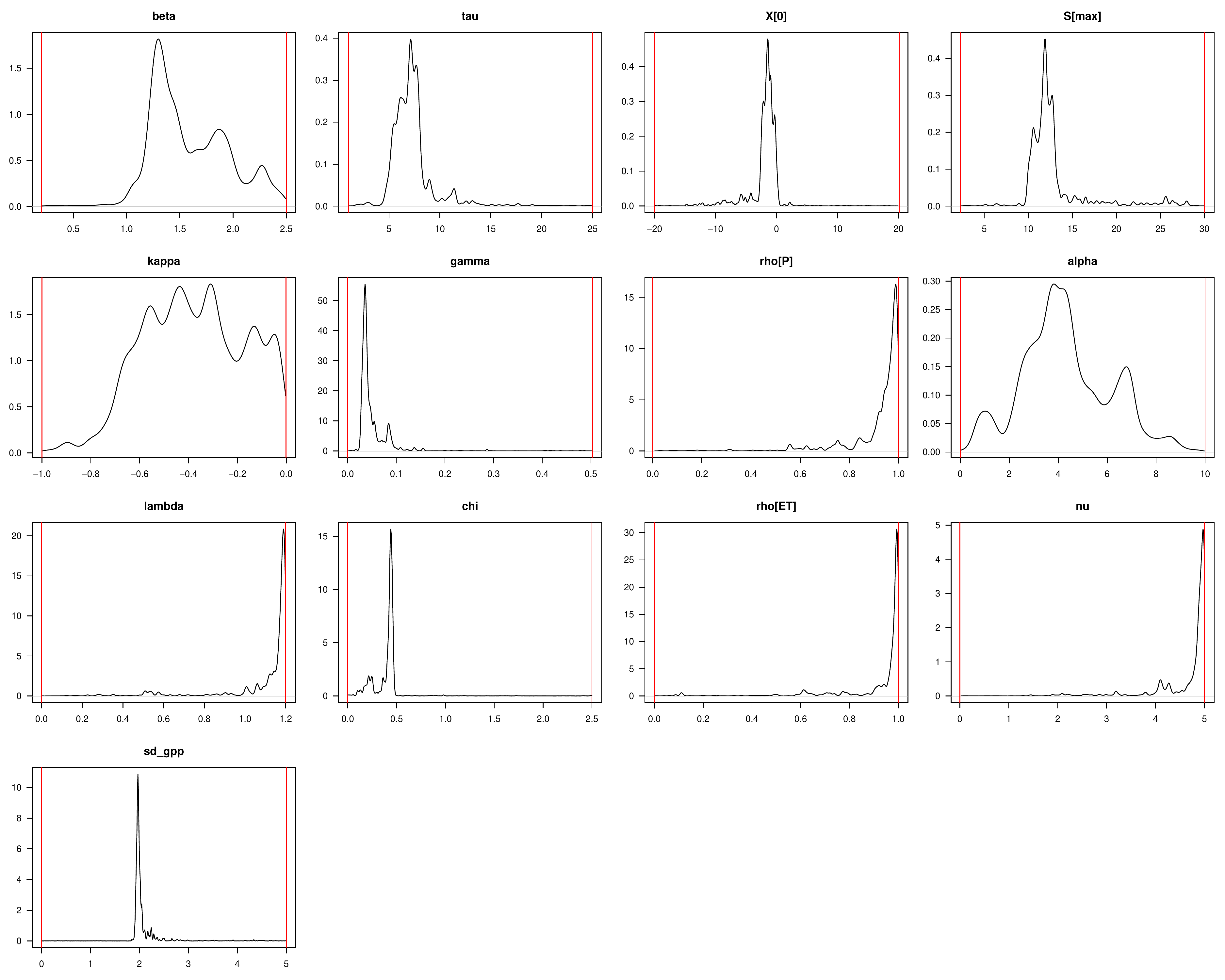}
    \caption{Posterior distributions of the re-calibrated PRELES parameters.}
    \label{fig:BayesianPosteriors}
\end{figure}

PRELES was re-calibrated using the Bayesian Tools package (see main document). A Markov Chain Monte Carlo simulation using the DREAMzs sampler was used. We ran three chains at 50000 iterations each. The results are shown in Fig. \ref{fig:BayesianPosteriors}.

\renewcommand{\bibname}{References}
\bibliographystyle{mybib.bst}
\bibliography{SI}